\pdfoutput=1
\documentclass[letterpaper,titlepage,11pt]{article}

\usepackage{amssymb,amsmath,amsfonts,jheppub}
\usepackage[toc]{appendix}
\numberwithin{equation}{section}
\usepackage[english]{babel}
\usepackage{dsfont}
\usepackage{amsmath,amssymb}
\usepackage{verbatim}
\usepackage{graphicx}
\usepackage{color}
\usepackage{amsmath}
\usepackage{amssymb}
\usepackage[utf8]{inputenc}

\DeclareFontFamily{OT1}{rsfs}{}
\DeclareFontShape{OT1}{rsfs}{m}{n}{ <-7> rsfs5 <7-10> rsfs7 <10->rsfs10}{}
\DeclareMathAlphabet{\mycal}{OT1}{rsfs}{m}{n}
\newcommand{\scri}{{\mycal I}}

\newcommand{\D}{\Delta}

\newcommand{\cA}{{\cal A}}
\newcommand{\cB}{{\cal B}}
\newcommand{\cC}{{\cal C}}
\newcommand{\cD}{{\cal D}}

\newcommand{\cF}{{\cal F}}
\newcommand{\cG}{{\cal G}}

\newcommand{\cL}{{\cal L}}

\newcommand{\cN}{{\cal N}}
\newcommand{\cO}{{\cal O}}

\newcommand{\bb}{\bar\beta}

\newcommand{\bt}{\tilde{\beta}}

\def\id{\mathds{1}}

\newcommand{\ket}[1]{|#1\rangle}
\newcommand{\bra}[1]{\langle#1|}
\newcommand{\bracket}[2]{\langle#1|#2\rangle}
\newcommand{\vev}[1]{\langle#1\rangle}

\def\one{{\hbox{ 1\kern-.8mm l}}}
\newcommand{\Dslash}{\not{\hbox{\kern-4pt $D$}}}
\newcommand{\pdslash}{\not{\hbox{\kern-2pt $\partial$}}}

\newcommand{\be}{\begin{equation}}
\newcommand{\ee}{\end{equation}}
\newcommand{\bea}{\begin{eqnarray}}
\newcommand{\eea}{\end{eqnarray}}
\newcommand{\ba}{\begin{array}}
\newcommand{\ea}{\end{array}}

\def\bbox{{\,\lower0.9pt\vbox{\hrule \hbox{\vrule height 0.2 cm
\hskip 0.2 cm \vrule height 0.2 cm}\hrule}\,}}
\newcommand{\dsl}{\pa \kern-0.5em /}

\newcommand{\ct}{\tilde{c}}
\newcommand{\thetat}{\tilde{\theta}}
\newcommand{\chit}{\tilde{\chi}}

\font\mybb=msbm10 at 12pt
\def\bb#1{\hbox{\mybb#1}}
\def\bZ {\bb{Z}}

\newcommand{\st}{\mathbf{t}}
\newcommand{\sx}{\mathbf{x}}

\newcommand{\mytitle}{Supersymmetric Galilean conformal blocks}

\title{\mytitle}

\newcommand{\il}{\ast}
\newcommand{\wm}{\dagger}
\newcommand{\zm}{\ddagger}

\author[\il]{Ivano Lodato,}
\emailAdd{ilodato@fudan.edu.cn}
\author[\wm]{Wout Merbis,}
\emailAdd{wmerbis@ulb.ac.be}
\author[\zm]{and Zodinmawia}
\emailAdd{zodin@iitk.ac.in}
\affiliation[\il]{Department of Physics and Center for Field Theory and Particle Physics, Fudan University, 220 Handan Road, 200433 Shanghai, China}
\affiliation[\wm]{Universit\'e Libre de Bruxelles and International Solvay Institutes, Physique Th\'eorique et Math\'ematique, Campus Plaine - CP 231, B-1050 Bruxelles, Belgium}
\affiliation[\zm]{Indian Institute of Technology Kanpur, Kalyanpur, Kanpur 208016, India}

\abstract{
We set up the bootstrap procedure for supersymmetric Galilean Conformal (SGC) field theories in two dimensions by constructing the SGC blocks in the $\cN=1$ and two possible $\cN =2$ extensions of the Galilean conformal algebra. In all analyzed cases, we present the bootstrap equations by crossing symmetry of the four point function.
In addition, we compute the global SGC blocks analytically by solving the differential equations obtained by acting with the Casimirs of the global subalgebras inside the four point function. These global blocks agree with the general SGC blocks in the limit of large central charge. 
We comment on possible applications to supersymmetric BMS$_3$ invariant field theories and flat holography.
}

\keywords{Supersymmetry, GCA symmetries, conformal blocks, conformal bootstrap, GCA bootstrap, BMS bootstrap}

\begin{document}

\maketitle

\section{Introduction}\label{se:1}

Conformal invariance is the most restrictive symmetry one can impose on a given relativistic (Lorentz invariant) quantum field theory \cite{Polyakov:1970xd}. In the absence of a physical scale the relevant physics is determined universally and is fixed by the different possible realizations of scale invariance. 
A systematic and non-perturbative method to find out which realizations of conformal symmetry are possible exists in the bootstrap program \cite{Ferrara:1973yt,Polyakov:1974gs}. The idea is to use unitarity, associativity of the operator algebra and crossing symmetry to constrain the possible sets of consistent conformal field theories (CFTs). This program has been successfully applied to minimal models in 2 dimensions and the (supersymmetric) Liouville model \cite{Belavin:1984vu,Zamolodchikov:1995aa,Belavin:2007gz}. A revival of this approach followed \cite{Dolan:2000ut,Dolan:2001tt} and the development of new techniques to constrain the space of possible CFTs in higher dimensions \cite{Rattazzi:2008pe} and solve the 3D Ising model \cite{ElShowk:2012ht} (see \cite{Rychkov:2016iqz,Simmons-Duffin:2016gjk,Poland:2018epd} for modern reviews and references). 

In the context of the AdS$_3$/CFT$_2$ correspondence, bootstrap techniques for CFTs with large central charge $c$ can provide valuable insights into quantum gravitational effects \cite{Hellerman:2009bu,Heemskerk:2009pn,Hartman:2014oaa,Fitzpatrick:2014vua,Jackson:2014nla,Chang:2015qfa,Lin:2015wcg,Lin:2016gcl,Benjamin:2016fhe,Chang:2016ftb}. The conformal blocks, which are the atomic constituents of CFT correlators and the basic ingredient for the conformal bootstrap, have a gravitational interpretation as the sum over exchanges of wave functions of primary objects in AdS$_3$, including AdS$_3$ gravitons. A detailed understanding of the blocks beyond the semi-classical limit would hence provide a way to sum graviton scattering processes \cite{Fitzpatrick:2015zha,Fitzpatrick:2015dlt,Fitzpatrick:2016mtp}. 

The bootstrap philosophy is not unique to relativistic CFTs. The same arguments lead to constraints on theories which possess a certain version of scale invariance, but are not invariant under (the full set of) Lorentz transformations. Examples of these kind of theories are the non- and ultra relativistic limits of CFTs \cite{Bagchi2009,Duval:2009vt,Bagchi2010,Bagchi:2010eg,Bagchi2012} and warped conformal field theories (WCFTs) \cite{Hofman:2011zj,Detournay:2012pc}, for which the bootstrap program has been formulated in \cite{Bagchi:2016geg,Bagchi:2017cpu} and \cite{Song:2017czq,Apolo:2018eky} respectively. In this work we will be interested in furthering the understanding of non- and ultra-relativistic limits of supersymmetric conformal field theories in two dimensions by computing the analogue of the conformal blocks in this class of theories. 

The motivation for considering (supersymmetric) non- and ultra-relativistic conformal field theories is fourfold. 
First of all, non-relativistic Galilean (super)conformal symmetries arise on the worldsheet of (super)string theory in the tensionless limit \cite{Bagchi2013a,Bagchi:2015nca,Bagchi:2016yyf,Bagchi:2017cte}. 
Secondly, just like conformal invariance appears at the fixed point in the RG flow of relativistic QFTs, Galilean conformal invariance is expected to arise at the fixed point for non-relativistic quantum field theories (or for effective field theories when velocities are small compared to the speed of light). A thorough understanding of Galilean conformal QFTs is therefore crucial to describe universal behavior in non-relativistic quantum field theories.
A third motivation is that very little explicit theories possessing this symmetry are known. Besides the aforementioned worldsheet theories of the tensionless string, we are only aware of the free scalar theory of ref. \cite{Barnich:2013yka}, a free fermion (from the supersymmetric generalization of the scalar) \cite{Barnich:2015sca} and an ultra-relativistic version of Liouville theory \cite{1210.0731}. It is conceivable that a successful implementation of the non- and ultra-relativistic conformal bootstrap program may lead to novel theories possessing these symmetries. 
Last but not least, our work is motivated by possible applications to a flat space holographic correspondence in 2+1 dimensions. We will elaborate on this a bit further.

The reason why the non- and ultra-relativistic limits of CFTs can to some extend be considered on the same footing is that in two spacetime dimensions the two symmetry algebras are isomorphic. The non-relativistic limit of the Virasoro algebra is given as an \.{I}n\"on\"u-Wigner contraction leading to the two dimensional Galilean conformal algebra (GCA$_2$). The ultra-relativistic limit is defined as a different contraction, but it leads to the same algebra. It is this limit which is relevant in the context of flat space holography because it gives the asymptotic symmetry algebra of flat space, the analogue of the Bondi-Metzner-Sachs algebra \cite{Bondi:1962,Sachs:1962wk} in three dimensions; the BMS$_3$ algebra \cite{Ashtekar:1996cd,Barnich:2006av}. 

The isomorphism between the non- and ultra-relativistic limit of the conformal algebra in two dimensions is perhaps not surprising since in two dimensions swapping time with the one spatial direction interchanges non- and ultra-relativistic physics. However for an implementation of the symmetry at the quantum level representations become important. In that sense the differences between GCA$_2$ and BMS$_3$ could become manifest given that the non- and ultra-relativistic limits of highest-weight representations of the Virasoro algebra are not the same \cite{Campoleoni:2016vsh}. Specifically, the non-relativistic limit leads to non-unitary highest-weight representations of GCA$_2$ while the ultra-relativistic limit instead leads to the unitary induced representations of BMS$_3$ \cite{Barnich:2014kra}. 

Contrary to what one could expect, the non-unitary highest-weight representation have proven very effective in flat space holography computations. When using these representations, one can match the entropy of flat space cosmological solutions with a modified Cardy formula \cite{Barnich2012a,Bagchi2013b}, stress-tensor correlators \cite{Fareghbal:2013ifa,Bagchi:2015wna} and entanglement entropy \cite{Bagchi2015,Jiang:2017ecm}. Recently it was shown that also the two and three point functions of generic GCA primaries, the Poincar\'e blocks \cite{Hijano:2017eii} and the BMS$_3$ blocks \cite{Hijano:2018nhq} match with geodesic Witten diagrams in asymptotically flat spacetimes. Using induced representations the only result known to us are the characters of BMS$_3$ \cite{Oblak:2015sea} but besides this it is not known how to compute correlators, entanglement entropy or the BMS analogue of the conformal blocks. So here we will keep using highest-weight representations and refer to the algebra as `Galilean conformal' instead of BMS. Our main results, however, are expected to also be applicable to BMS$_3$ in the highest-weight representations by the above considerations. The interesting question on how to reproduce these results in flat holography using induced (or any other unitary) representation of BMS$_3$ we will leave for future work.

In this paper, we will compute the non-relativistic version of superconformal blocks \cite{Kiritsis:1987np,AlvarezGaume:1991bj,Belavin:2006zr,Belavin:2007gz,Belavin:2007eq,Hadasz:2006qb,Hadasz:2008dt,Fitzpatrick:2014oza,Chen:2016cms,Cornagliotto:2017dup,Alkalaev:2018qaz,Hikida:2018eih} or the supersymmetric Galilean conformal (SGC) blocks. Various supersymmetric extensions of GCA$_2$/BMS$_3$ symmetries and gravitational bulk theories with these asymptotic symmetries have been considered in \cite{Awada:1985by,Bagchi:2009ke,Mandal:2010gx,Banks:2014iha,Barnich:2014cwa,Banerjee:2015kcx,Lodato:2016alv,Banerjee:2016nio,Banerjee:2017gzj,Basu:2017aqn,Fuentealba:2017fck}. Here we will focus on setting up the bootstrap procedure in the $\cN=1$ and two possible $\cN=2$ supersymmetric extensions which we define from a contraction of the $\cN=(1,1)$ superconformal algebra in the next subsection. In each of these cases we will find explicit expressions for the global blocks (for light operators in the limit of large central charges) and formulate the bootstrap equations.

\subsection{Non- and ultra-relativistic superconformal algebras}
Before we begin with the intrinsic analysis we recall how the two $\cN=2$ symmetry algebras we consider in this paper follow from non- and ultra-relativistic limits of the $\cN=(1,1)$ superconformal algebra \cite{Mandal:2010gx,Banerjee:2016nio}. The superconformal algebra consists out of two super Virasoro algebras
\begin{subequations}
\begin{align}
	[\mathcal{L}^\pm_n, \mathcal{L}^\pm_m] &  = (n-m) \mathcal{L}^\pm_{n+m} + \tfrac{c^\pm}{12} n(n^2-1) \delta_{n+m,0} \,,  \\
	[\mathcal{L}^\pm_n, \mathcal{G}^\pm_r] & = (\tfrac{n}{2}-r) \mathcal{G}^\pm_{n+r} \,, \\ 
	\{ \cG^\pm_r, \cG^\pm_s \} & = 2 \cL_{r+s}^{\pm} + \tfrac{c^\pm}{3}(r^2 - \tfrac{1}{4})\delta_{r+s,0}\,.
\end{align}
\end{subequations}
For the bosonic part of the algebra we have two different ways to contract the generators, one corresponding to the non-relativistic limit and one to the ultra-relativistic limit \cite{Barnich:2006av,Bagchi2012}. They are defined by taking $\epsilon \to 0$ after redefining the Virasoro generators as:
\begin{subequations}
	\label{contractions}
\begin{align}
 \text{non-relativistic:} &&& 	L_n = \mathcal{L}^+_n + \mathcal{L}^-_n \,, \quad M_n = \epsilon (\mathcal{L}^+_n - \mathcal{L}^-_n )\,,		\\
 \text{ultra-relativistic:} &&& L_n = \mathcal{L}^+_n - \mathcal{L}^-_{-n} \,, \quad M_n = \epsilon (\mathcal{L}^+_n + \mathcal{L}^-_{-n} )\,.
\end{align}
\end{subequations}
Note that in the ultra-relativistic limit the raising and lowering operators mix in the contraction. This is the reason why the highest-weight representations of Virasoro do not remain highest-weight in this limit. Instead they become massive rest-frame states which are annihilated by all $M_n$ with $n\neq0$. Acting with any $L_n$ for $n\neq0$ builds up the BMS module of \cite{Campoleoni:2016vsh}. 

After taking $\epsilon \to 0$ the bosonic part reduces to the GCA$_2$/BMS$_3$ algebra
\begin{subequations}
	\label{GCA2}
	\begin{align}
	[L_n, L_m] & = (n-m)L_{m+n} + \frac{c_L}{12} n (n^2-1) \delta_{m+n,0}\,, \\
	[L_n, M_m] & = (n-m)M_{m+n} + \frac{c_M}{12}n (n^2-1) \delta_{m+n,0}\,, \\
	[M_n, M_m] & = 0\,,
	\end{align}
\end{subequations}
with central charges $c_M = \epsilon (c^+ - c^-)$ and $c_L = c^+ + c^-$ in the non-relativistic limit and $c_M = \epsilon (c^+ + c^-)$ and $c_L = c^+ - c^-$ for the ultra-relativistic case.\footnote{Equivalently the \.{I}n\"on\"u-Wigner contraction can also be implemented by taking $\epsilon = 1/\ell $ to be a Grassmann valued parameter \cite{Krishnan:2013wta}}
The difference between GCA$_2$ and BMS$_3$ exists in how the generators act as spacetime symmetries. In the case of GCA$_2$ we are considering two dimensional field theories invariant under time-dependent accelerations $g(t)$ and time reparametrizations $f(t)$ acting on the coordinates as
\begin{equation}\label{GCAcoord}
t \to f(t) \,, \qquad \qquad x \to \partial_t f(t) x + g(t)\,.
\end{equation}
These transformations are generated by the vector fields on the Galilean plane $\mathbb{R}^{1,1}$.
\begin{subequations}
	\label{GCAgenerators}
	\begin{align}
	L_n & = - t^{n+1} \partial_t - (n+1)x t^n \partial_x\,, \\
	M_n & = - t^{n+1}\partial_x\,.
	\end{align}
\end{subequations}
These vector fields span the GCA$_2$ algebra \eqref{GCA2}, which allows for two non-trivial central extensions, denoted here as $c_L$ and $c_M$.

The global subgroup of this algebra consists out of spatial translations ($M_{-1} = -\partial_x$), time translations ($L_{-1}=-\partial_t$), dilations ($L_0=-(t\partial_t + x \partial_x)$), Galilean boosts ($M_0=-t \partial_x$), and a non-relativistic version of the spatial and temporal special conformal transformations ($M_{+1}=-t^2 \partial_x$) and ($L_{+1}= -t^2\partial_t - 2xt\partial_x$) respectively. 

For BMS$_3$ we interpret $M_n, L_n$ in \eqref{GCA2} as generating supertranslations and superrotations along null infinity ($\scri$):
\begin{equation}
x \to f(x) \,, \qquad \qquad u \to \partial_x f(x) u + g(x)\,.
\end{equation}
Here $\{u,x\}$ are coordinates on a null plane. The null cylinder coordinates along $\scri$, more customary for flat space holography, ($\tau, \varphi)$ can be obtained by the map $x = e^{i\varphi}, u = i \tau e^{i\varphi}$. The map between representation independent results in GCA and BMS exists in swapping the time coordinate $t$ with the spatial coordinate at $\scri$ and exchanging $x$ in \eqref{GCAcoord} with the null direction along $\scri$. 

In both the relativistic and the non-relativistic limit the fermionic generators can be taken to scale in two different ways. One option is to scale both $\cG^{\pm}_r$ in the same way, which leads to the democratic (or homogeneous) $\cN=2$ super GCA algebra. The second option is to take the fermionic generators to scale as an asymmetric combination, much like the bosonic generators in \eqref{contractions}. We will refer to this as the despotic algebra, after \cite{Lodato:2016alv}.
\begin{align}
& \text{Democratic} && \text{Despotic} \nonumber \\
\text{non-relativistic}: \qquad & \sqrt{2} Q^{\pm}_r = \sqrt{\pm \epsilon} \, \cG_r^{\pm} \,, &&  G_r = \cG^+_r + 
\cG^-_r \,, \\
& && H_r = \epsilon (\cG_r^+ - \cG_r^- )\,, \nonumber \\
\text{ultra-relativistic}: \qquad & \sqrt{2} Q^{\pm}_r = \sqrt{\epsilon} \,  \cG_{\pm r}^{\pm}\,, && G_r = \cG_{r}^+ - i \cG_{-r}^-\,,  \\
& && H_r = \epsilon (\cG_r^+ + i \cG_{-r}^- )\,. \nonumber
\end{align}
The democratic limit in both cases leads to \eqref{GCA2} together with
\begin{subequations}
	\label{SGCAdemocratic1}
	\begin{align}
	[L_n,Q^{\pm}_{r}] & = (\tfrac{n}{2}-r)Q^{\pm}_{r+n}\,, \\
	\{Q^{\pm}_{r}, Q^{\pm}_{s} \} & = M_{r+s} + \frac{c_M}{6} (r^2-\tfrac14)  \delta_{r+s,0}\,, \\
	\{Q^{\pm}_{r},Q^{\mp}_{s} \} &  = 0 = [M_n, M_m] = [M_n, Q^{\pm}_{r}] \,,
	\end{align}
\end{subequations}
while the despotic limit gives the (anti)-commutators
\begin{subequations}
	\label{SGCAdespotic1}
	\begin{align}
	[L_n,G_{r}] & = (\tfrac{n}{2}-r)G_{r+n}\,,
	\\
	[L_n,H_r]&=[M_n,G_r]= (\tfrac{n}{2}-r)H_{r+n}\,, \\
	\{G_{r}, G_{s} \} & = 2\,L_{r+s} + \frac{c_L}{6} (r^2-\tfrac14)  \delta_{r+s,0}\,, \\
	\{G_r,H_s\} &= 2\,M_{r+s}+ \frac{c_M}{6} (r^2-\tfrac14)  \delta_{r+s,0}\,,
	\\
	[M_n, M_m]&= 0 = [M_n, H_{r}]=\{H_r,H_s\} \,.
	\end{align}
\end{subequations}
Both supersymmetry algebras follow as asymptotic symmetries of flat space supergravities in three dimensions \cite{Lodato:2016alv}. A subtlety is that the limit of the Hermitian conjugate is only well defined in the ultra-relativistic democratic case and in the non-relativistic despotic case. 

In the remainder of this paper we will work intrinsically with field theories invariant under the algebras \eqref{SGCAdemocratic1} and \eqref{SGCAdespotic1}. The details on the various limits we will leave for future work \cite{toappear}.
In the next section we will review the bosonic case first derived in \cite{Bagchi:2016geg} and fix notation for the remainder of the paper. 
In section \ref{se:3} we will formulate the Galilean conformal bootstrap and compute the global blocks for the $\cN=1$ super-GCA algebra, which is essentially \eqref{SGCAdemocratic1} with only one set of supergenerators.
In section \ref{se:4} we will discuss the same in the $\cN=2$ democratic case and section \ref{se:5} takes care of the despotic case. 
We present our conclusions in section \ref{se:conc}.
Appendix \ref{sec:OPE} collects various details about the operator product expansions for these theories.

\section{Galilean conformal blocks: a review}\label{se:2}
In this section we will fix the notation and warm up with a brief review of the GCA invariant field theories and the GCA$_2$ (or BMS$_3$)-blocks constructed in \cite{Bagchi:2016geg}. A thorough account of the setting, definitions and computations here can be found in \cite{Bagchi2009,Bagchi2010,Bagchi:2017cpu} to which we refer for more details.

\subsection{A GCA state-operator correspondence}
The basic idea is to consider possible correlation functions of operators in QFTs invariant under GCA$_2$ symmetry \eqref{GCA2}. Like in conformal field theories, it will turn out to be very useful to define a state-operator correspondence in the highest-weight representations of GCA$_2$. The vacuum state $\ket{0}$ is defined as being annihilated by the global subgroup and all lowering operators
\begin{equation}
\{L_n,M_n\}|0\rangle = 0 \,,\qquad \forall\, n \geq -1\,.
\end{equation}
Primary states are labeled by their $L_0$ and $M_0$ weights, denoted by $\Delta, \xi$ respectively:
\begin{equation}
L_0 |\Delta, \xi \rangle = \Delta | \Delta, \xi \rangle\,, \qquad M_0 |\Delta, \xi \rangle = \xi | \Delta, \xi \rangle \,.
\end{equation}
They are annihilated by lowering operators $\{L_n, M_n\}$ with $n>0$ and they correspond to Galilean conformal primary operators $\phi_p(t,x)$ inserted at the origin of the Galilean plane $\mathbb{R}^{1,1}$
\begin{equation}
\phi_p(0,0)\ket{0} \equiv \ket{\Delta_p,\xi_p}\,.
\end{equation} 
The GCA modules (analogue to the Verma modules in CFT) are then defined as acting with raising operators $\{L_n, M_n\}$, with $n<0$ on the primary states, defining the GCA descendant states at level $N$
\begin{equation}\label{descendants}
\ket{\Delta,\xi,\{N\}} = L_{-\{k\}} M_{-\{l\}} \ket{\Delta,\xi} \equiv L_{-k_1} \ldots L_{-k_i} M_{-l_1} \ldots M_{-l_j} \ket{\Delta,\xi}\,,
\end{equation}
where $\{N\}$ denotes two sets of integers $\{ k\} $ and $\{l\}$, whose total level $N$ is the sum of all elements in the sets and we organize them in descending order.
The next ingredients we need are the definition of the out state and the Hermitian conjugate. The Hermitian conjugate of a primary state can be represented as an operator in the infinite future
\begin{equation}\label{out}
\bra{\Delta_p, \xi_p} = \lim_{t \to \infty} t^{2\Delta_p} \bra{0} \phi_p(t,0)\,.
\end{equation}
Hermitian conjugation inverts the order of the descendant operators and takes
\begin{equation}
L_{k}^\dagger = L_{-k} \,, \qquad M_l^\dagger = M_{-l}\,.
\end{equation}
Hence
\begin{equation}
\bra{\Delta, \xi, \{N\}} = \bra{\Delta,\xi}M_{l_j} \ldots M_{l_1}L_{k_i}\ldots L_{k_1} \,.
\end{equation}
The out states \eqref{out} are annihilated by the raising operators.

The primary operators $\phi_p(t,x)$ transform under GCA symmetries as \cite{Bagchi2010}:
\begin{subequations}
\label{GCAcommutators}
\begin{align}
\delta_{L_n}\phi_p(t,x) = [L_n, \phi_p(t,x)] & = \big[t^{n+1} \partial_t + (n+1)x t^n \partial_x + \xi_p n(n+1)x t^{n-1} \\
& \qquad \nonumber + \Delta_p (n+1)t^n \big] \phi_p(t,x) \;, \\
\delta_{M_n}\phi_p(t,x) = [M_n, \phi_p(t,x)] & = \big[ t^{n+1}\partial_x + \xi_p (n+1) t^n \big] \phi_p(t,x)\,. 
\end{align}
\end{subequations}
We will denote the right hand side of these equations as differential operators $\cD_{L_n}$ and $\cD_{M_n}$. Applying these operators inside correlation function gives the GCA Ward identities.

The correlation functions between primaries are invariant under the global Galilean conformal algebra. This fixes the functional form of the normalized two-point function completely 
\begin{equation}\label{twopt}
\langle \phi_m(t_m,x_m) \phi_n(t_n,x_n) \rangle = \frac{\delta_{\Delta_m,\Delta_n} \delta_{\xi_m,\xi_n}}{t_{mn}^{\Delta_m+\Delta_n}}e^{ - \frac{x_{mn} (\xi_m + \xi_n)}{t_{mn}}}\,,
\end{equation}
where $t_{mn} = t_m - t_n$ and likewise for $x_{mn}$. The three point function between primaries depend on a single coefficient $c_{imn}$
\begin{equation}\label{threept}
\langle \phi_i(t_i,x_i) \phi_m(t_m,x_m) \phi_n(t_n,x_n) \rangle = \frac{c_{imn}}{t_{im}^{\Delta_{imn}} t_{mn}^{\Delta_{mni}} t_{in}^{\Delta_{inm}}}e^{-\frac{x_{im} \xi_{imn}}{t_{im}}-\frac{x_{mn} \xi_{mni}}{t_{mn}}-\frac{x_{in} \xi_{inm}}{t_{in}} }\,,
\end{equation}
where $\Delta_{imn} = \Delta_i + \Delta_m - \Delta_n$ and likewise for $\xi_{imn}$. 
The four point function can depend on a general function of the invariant cross ratios $T$ and $X$. We will write it as
\begin{align}\label{fourpt}
& \langle \phi_i(t_i,x_i) \phi_j(t_j,x_j) \phi_m(t_m,x_m) \phi_n(t_n,x_n) \rangle = t_{ij}^{-2\Delta_{i}} t_{jm}^{\Delta_i -\Delta_j -\Delta_m + \Delta_n} t_{jn}^{\Delta_i - \Delta_j + \Delta_m -\Delta_n}   \\
& \times t_{mn}^{-\Delta_{i} + \Delta_j -\Delta_m - \Delta_n} e^{-\frac{2x_{ij} \xi_{i}}{t_{ij}}+\frac{x_{jm}(\xi_i -\xi_j - \xi_m +\xi_n)}{t_{jm}} + \frac{x_{jn}(\xi_i -\xi_j + \xi_m -\xi_n)}{t_{jn}} + \frac{x_{mn} (-\xi_i+\xi_j - \xi_m - \xi_n)}{t_{mn}}} F_{\rm GCA}(T,X)\,, \nonumber
\end{align}
where $F_{\rm GCA}$ is an arbitrary function of the GCA cross ratios
\begin{equation}\label{GCAcrossratios}
T = \frac{t_{ij} t_{mn}}{t_{im}t_{jn}}\,, \qquad \frac{X}{T} = \frac{x_{ij}}{t_{ij}} + \frac{x_{mn}}{t_{mn}} - \frac{x_{im}}{t_{im}} - \frac{x_{jn}}{t_{jn}} \,.
\end{equation}
We have chosen conventions for the four point function such that at equal external weights $\Delta_{i,j,m,n} = \Delta$ and $\xi_{i,j,m,n} = \xi$ the prefactor simplifies to
\begin{equation}
\langle \phi(t_i,x_i) \phi(t_j,x_j) \phi(t_m,x_m) \phi(t_n,x_n) \rangle = \frac{1}{(t_{ij}t_{mn})^{2\Delta}} e^{-\frac{2x_{ij} \xi}{t_{ij}} - \frac{2x_{mn}\xi}{t_{mn}}} F_{\rm GCA}(T,X)\,.
\end{equation}
By a global GCA transformation we can always take the coordinates to the special values
\begin{equation}\label{points}
\{ (t_i,x_i), (t_j,x_j), (t_m,x_m), (t_n,x_n)\} = \{(\infty,0), (1,0), (t,x), (0,0) \}\,,
\end{equation}
under which $T \to t$ and $X \to x$ and the correlation functions \eqref{twopt}, \eqref{threept} and \eqref{fourpt} become
\begin{subequations}
\label{correlators}
\begin{align}
\langle \Delta_i, \xi_i | \Delta_n, \xi_n \rangle & \equiv \lim_{t_i \to \infty} t_i^{2\Delta_i}\langle \phi_i(t_i,x_i) \phi_n(0,0)\rangle = \delta_{\Delta_i,\Delta_n}\delta_{\xi_i,\xi_n}\,, \\ 
\langle \Delta_i, \xi_i |\phi_j(1,0) | \Delta_n, \xi_n \rangle & \equiv \lim_{t_i \to \infty} t_i^{2\Delta_i}\langle \phi_i(t_i,x_i) \phi_j (1,0) \phi_n(0,0)\rangle = c_{ijn}\,, \\
\langle \Delta_i, \xi_i |\phi_j(1,0)\phi_m(t,x)| \Delta_n, \xi_n \rangle & \equiv \lim_{t_i \to \infty} t_i^{2\Delta_i}\langle \phi_i(t_i,0)\phi_j(1,0)\phi_m(t,x) \phi_n(0,0)\rangle  \\
 = & \frac{F_{\rm GCA}(t,x)}{t^{\Delta_i-\Delta_j+\Delta_m +\Delta_n} (1-t)^{-\Delta_i+\Delta_j+\Delta_m-\Delta_n}} e^{-\frac{x(\xi_i-\xi_j+(1-2t)\xi_m + \xi_n)}{t(1-t)}}\,. \nonumber
\end{align}
\end{subequations}
The matrix of inner products of states including descendants defines the GCA analogue of the Kac matrix in CFTs. We will denote it by $\mathfrak{M}_N$ and its matrix elements are
\begin{equation}\label{Kacmat}
\mathfrak{M}_{\{N'\},\{N\}} = \bracket{\Delta,\xi,\{N'\}}{\Delta,\xi, \{N\}} . 
\end{equation}
It is orthogonal in the sense that only states at the same level have non-zero Kac matrix elements. At level one we have
\begin{equation}
\mathfrak{M}_1 = \left[ \begin{array}{c} \bra{\Delta,\xi} L_{+1} \\ \bra{\Delta,\xi} M_{+1} \end{array} \right] \bigg[ L_{-1}\ket{\Delta,\xi} \; M_{-1} \ket{\Delta,\xi} \bigg] = \left( \begin{array}{cc} 2 \Delta & 2\xi \\ 2\xi & 0 \end{array}\right) \,.
\end{equation}
From this one can easily see that $\det(\mathfrak{M}_1) = -4\xi^2$ is negative, indicating the presence of negative norm states and hence a breakdown of unitarity in this representation. 

At level two we have
\begin{align}
\mathfrak{M}_2 = &\; \left[ \begin{array}{c}  \bra{\Delta,\xi} L_{+1}L_{+1} \\ \bra{\Delta,\xi} L_{+2} \\\bra{\Delta,\xi}M_{+1} L_{+1} \\ \bra{\Delta,\xi} M_{+2} \\ \bra{\Delta,\xi} M_{+1} M_{+1} \end{array} \right] \bigg[  L_{-1}L_{-1} \ket{\Delta,\xi} \: L_{-2}\ket{\Delta,\xi} \: M_{-1} L_{-1}\ket{\Delta,\xi} \: M_{-2} \ket{\Delta,\xi} \; M_{-1} M_{-1}\ket{\Delta,\xi}  \bigg] \nonumber \\
= &\; \left( \begin{array}{ccccc} 
4\Delta(1 + 2\Delta)	& 6\Delta 		& 4\xi(1+2\Delta) 	  & 6\xi	& 8\xi^2 		\\ 
6\Delta			& \frac{c_L}{2}+4\Delta	&  6\xi & \frac{c_M}{2}+4\xi		& 0	\\
 4\xi(1+2\Delta)	& 6\xi			& 4\xi^2 	  & 0			& 0		\\
 6 \xi			& \frac{c_M}{2}+4\xi	&0		  & 0			& 0		\\
 8\xi^2		& 0			& 0		  & 0			& 0
\end{array}\right) \,,
\end{align}
and $\det(\mathfrak{M}_2) = (2\xi)^6 (c_M + 8 \xi)^2 $. At level 3 the determinant of the Kac matrix is $\det(\mathfrak{M}_3) = - 36(2\xi)^{16} (c_M + 3\xi)^2(c_M + 8\xi)^4$. For $c_M=0$ and $\xi=0$ the Kac matrices are block diagonal, containing a chiral CFT part generated by $L_n$ and all states involving $M_n$ generators become null states. In that case the theory trivially reduces to a chiral CFT.

\subsection{The Galilean conformal blocks} \label{se:2.2}

The GCA block is defined analogously to the conformal blocks by expressing the four point function as a sum over primaries (labeled by $p$) in a given exchange channel. In this case we will choose the exchange channel $ji \leftrightarrow mn$ and define the Galilean conformal blocks $F^{ji}_{mn}(p|t,x)$ as the addends in the expansion:
\begin{equation}\label{blockdef}
G^{ji}_{mn}(t,x) \equiv \bra{\Delta_i,\xi_i}\phi_j(1,0) \phi_m(t,x) \ket{\Delta_n,\xi_n} = \sum_p c^p_{ji}c^p_{mn}F^{ji}_{mn}(p|t,x)\,.
\end{equation}
We may compute it as a series expansion of products of three point functions by inserting a complete basis of states $\id = \sum_{\alpha} \ket{\alpha}\bra{\alpha}$ in the middle of the four point function \eqref{blockdef}. In the highest-weight representation the complete basis of states is given by
\begin{equation}\label{basis}
\id = \sum_{p,\{N'\},\{N\}} \ket{\Delta_p, \xi_p, \{N'\}} \mathfrak{M}^{\{N'\},\{N\}}\bra{\Delta_p,\xi_p,\{N\}}\,.
\end{equation}
Here $\mathfrak{M}^{\{N\},\{N'\}}$ (with indices up) is the inverse of the GCA Kac matrix \eqref{Kacmat}. The blocks can now be written as
\begin{align}\label{GCAblock}
F^{ji}_{mn}(p|t,x) = (c^p_{ji})^{-1} (c^p_{mn})^{-1} \sum_{\{N\},\{N'\}} \bra{\Delta_i, \xi_i} & \phi_j(1,0) \ket{\Delta_p,\xi_p, \{N'\}}  \mathfrak{M}^{\{N'\},\{N\}} \\
& \nonumber \times \bra{\Delta_p,\xi_p,\{N\}} \phi_m(t,x) \ket{\Delta_n,\xi_n} \,. 
\end{align}
The three point functions involving descendants in this expression can be computed by acting with the relevant differential operators \eqref{GCAcommutators} on the three point function involving only primary fields
\begin{align}
\bra{\Delta_p,\xi_p,\{N\}} \phi_m(t,x) \ket{\Delta_n,\xi_n} & = 
\bra{\Delta_p,\xi_p} [M_{l_j} ,\ldots [ M_{l_1} ,[L_{k_i}, \ldots, [L_{k_1}, \phi_m(t,x)]]]] \ket{\Delta_n,\xi_n} \nonumber \\
& = \cD_{L_{k_1}} \ldots \cD_{L_{k_i}} \cD_{M_{l_1}} \ldots \cD_{M_{l_j}} e^{-\xi_{mnp}\frac{x}{t}}t^{-\Delta_{mnp}} \,.
\end{align}
Here in the last line we have used that $\bra{\Delta_p,\xi_p} \phi_m(t,x) \ket{\Delta_n,\xi_n} = e^{-\xi_{mnp}\frac{x}{t}}t^{-\Delta_{mnp}} $. The first three point function in \eqref{GCAblock} is computed analogously, using generic coordinate dependence for $(t_j,x_j)$ and setting it to $(1,0)$ at the end. 

The first few terms of the sum \eqref{GCAblock} are given explicitly as
\begin{align}\label{blockexp}
F^{ji}_{mn}(p|t,x) =  e^{-\xi_{mnp}\frac{x}{t}}t^{-\Delta_{mnp}} & \bigg[1 + \frac{t}{2\xi_p} \left( \Delta_{pji}\xi_{pmn} + \xi_{pji}\Delta_{pmn} - \frac{\xi_{pji}\xi_{pmn} \Delta_p}{\xi_p} \right) \\
& \nonumber + x\, \frac{ \xi_{pji}\xi_{pmn}}{2\xi_p} + \ldots \bigg]\,.
\end{align}
Here the dots denote terms of level 2 and higher.
The resulting block can equivalently be expressed in terms of products of coefficients $\beta$ appearing in the operator product expansion (OPE)
\begin{equation}
F^{ji}_{mn}(p|t,x) =  \frac{ e^{-\xi_{mnp}\frac{x}{t}} }{t^{\Delta_{mnp}}}\sum_{\{N\},\{N'\},\alpha} t^{N-\alpha}x^\alpha  \beta_{ji}^{p,\{N'\},0}\, \mathfrak{M}_{ \{N'\}, \{N\} } \beta_{mn}^{p,\{N\},\alpha}\,.
\end{equation}
For a more detailed definition of the $\beta$ coefficients, as well as the definition of the OPE itself, we refer to appendix \ref{sec:OPE}.

Crossing symmetry implies the four point function \eqref{blockdef} should not depend on the exchange channel under consideration. Instead of computing the blocks in the channel $ji \leftrightarrow mn$, we may opt to compute them in a different channel, say $ni \leftrightarrow mj$. This effectively switches $j$ with $n$, which from the point of view of \eqref{GCAcrossratios} takes $t \to 1-t$ and $x \to -x$. Under this crossing symmetry transformation the GCA block should satisfy
\begin{equation}\label{GCAbootstrap}
\sum_p c^p_{ji} c^p_{mn} F^{ji}_{mn}(p|t,x) = \sum_{p'} c^{p'}_{ni}c^{p'}_{mj} F_{ni}^{mj}(p'|1-t,-x)\,.
\end{equation} 
This is the GCA bootstrap equation, first derived in \cite{Bagchi:2016geg}. One of our main results here is to derive the supersymmetric analogue of this equation.

\subsection{Large central charge expansion and the global GCA-block}

The bootstrap philosophy is to use equation \eqref{GCAbootstrap} to constrain the structure constants and weights of primary operators consistent with the (in this case GCA) symmetries of the theory. Even though the symmetry algebra is infinite dimensional and thus highly constraining, it is quite difficult to find a closed form for the GCA blocks in general. But like in CFTs, it is possible to derive a closed form expression for the GCA blocks in the limit of large central charge. We will briefly review how to obtain the GCA blocks in this limit, to which we will refer as global GCA blocks.

In the limit of large central charges the sum over the exchanged state in \eqref{basis} is dominated by primaries and descendants generated by $L_{-1}$ and $M_{-1}$.\footnote{One can prove this by considering the contributions of descendants generated by $L_{-s}$ and $M_{-s}$ with $s>1$ to the normalized identity operator $\id = \sum_{\alpha} \frac{\ket{\alpha} \bra{\alpha}}{ \bracket{\alpha}{\alpha} } $ in an orthogonal basis. Since the norm of these states are of order $(c_M/\xi, c_L)$, the contributions to $\id$ are subleading in the large central charge expansion. Descendant states with more $L_{-s}$ and $M_{-s}$ insertions are suppressed even further. Similar arguments hold for the supersymmetric cases discussed below.}
These descendants commute with the quadratic Casimirs of the global subalgebra of GCA$_2$. Inserting the Casimirs to the right of the identity operator in \eqref{GCAblock} we can derive a differential equation by commuting it through the operators on both sides. By commuting it through the descendants on the left we find only the eigenvalues of the Casimirs on the exchanged state $\bra{\Delta_p,\xi_p}$. When commuting the Casimir through the right, by means of \eqref{GCAcommutators} we find a differential operator acting on the four point function. This process results in a differential equation which we can solve.

We can define the global GCA block $g^{ji}_{mn}(p|t,x)$ as a limit of the GCA block:
\begin{equation}
g^{ji}_{mn}(p|t,x) = \lim_{c_{M,L} \to \infty} F^{ji}_{mn}(p|t,x) \,.
\end{equation}
In writing $c_{M,L} \to \infty$ we are slightly abusing the notation because the central charge $c_M$ is dimensionful and hence its value can always be rescaled to one. What we mean is that we only consider primary operators $\phi$ which are light, such that the dimensionless ratio $\xi/c_M \to 0$. 

The Casimirs of ISO(2,1) are
\begin{subequations}
\label{GCACasimirs}
\begin{align}
\cC_1 & = M^2_0 - M_{-1}M_{+1} \,, \\
\cC_2 & = 2 L_0 M_0 - \frac12 \left( M_{-1}L_{+1} + M_{+1}L_{-1} + L_{-1}M_{+1} + L_{+1}M_{-1} \right)\,.
\end{align}
\end{subequations}
By acting with the Casimirs inside the four point function as outlined above we can derive the eigenvalue equations
\begin{align}
& \bra{\Delta_i,\xi_i} \phi_j(1,0) \cC_{1,2} \phi_m(t_m,x_m)\phi_n(t_n,x_n) \ket{0}\big|_{ \{(t_m,x_m),(t_n,x_n)\} = \{(t,x),(0,0) \} }\nonumber \\
&  \qquad = \sum_p \bra{\Delta_i,\xi_i} \phi_j(1,0)\ket{\Delta_p,\xi_p,\{N'\}} \mathfrak{M}^{\{N'\},\{N\}} \bra{\Delta_p,\xi_p,\{N\}}\cC_{1,2} \phi_m(t,x)\ket{\Delta_n,\xi_n} \nonumber \\
&   \qquad  = \sum_p \lambda^p_{1,2} c^p_{ji}c^p_{mn} g^{ji}_{mn}(p|t,x) +  \cO \left( \frac{\xi_p}{c_M},\frac{1}{c_L} \right)\,.
\end{align}
Here in the first line it is understood that one first acts with the Casimirs on the primaries $\phi_m(t_m,x_m)$ and $\phi_n(t_n,x_n)$ inside the four point function \eqref{fourpt} as differential operators and takes the coordinates $(t_m,x_m)$ and $(t_n,x_n)$ to their special values \eqref{points} afterwards. In the last line the eigenvalues of the Casimirs on the primary in the exchange channel $\bra{\Delta_p,\xi_p}$ appear. They are:
\begin{equation}
\lambda^p_1 = \xi_p^2 \,,\qquad \qquad \lambda^p_2 = 2\xi_p(\Delta_p -1) \,.
\end{equation}
At equal external weights, $\Delta_{i,j,m,n} = \Delta $ and $ \xi_{i,j,m,n} = \xi$, the Casimir equations can be written as
\begin{align}
\cD_x^2 f(p|t,x)&  = 0 \,, && \cD_{tx} f(p|t,x) = 0\,,
\end{align}
with
\begin{equation}\label{DCas}
\cD_x^2 = \lambda_1^p - t^2(1-t)\partial_x^2\,, \qquad \text{and} \qquad \cD_{tx} =   2(1-t)t^2 \partial_t \partial_x + (2-3t) \partial_x^2 - 2t^2 \partial_x - \lambda_2^p \,.
\end{equation}
Here the function $f(p|t,x)$ is defined as the summand of the large $c$ limit of $F_{\rm GCA}$ appearing in \eqref{fourpt}; i.e.  $\sum_p c^p_{ji}c^p_{mn} f(p|t,x) = \lim_{c_{M,L}\to\infty} F_{\rm GCA}(t,x)$\,.

The solution to these equations involves two arbitrary integration constants which can be fixed by comparing the series expansion of the solution to (the large central charge limit of) the expansion of the GCA block \eqref{blockexp}. The final answer for equal external weights is
\begin{equation}\label{bosblock}
g_{\Delta,\xi}(p|t,x) = \frac{e^{-2\xi \frac{x}{t} } } {t^{2\Delta}} f(p|t,x) =  \frac{e^{ -2 \xi \frac{x}{t} + \xi_p \frac{x}{t\sqrt{1-t}} } }{t^{ 2 \Delta - \Delta_p } \sqrt{1-t} }  \left( \frac12 +\frac12 \sqrt{1-t} \right)^{2-2\Delta_p}\,,
\end{equation}
where $0<t<1$. 
In the coming sections, we will concern ourselves with computing the supersymmetric extensions of the global Galilean conformal block.

\section{$\cN=1$ supersymmetric Galilean conformal blocks}\label{se:3}

In this section we will discuss the Galilean conformal blocks in the simplest supersymmetric extension of the Galilean conformal algebra. This $\cN=1$ extension is isomorphic to the $\cN =1$ super BMS$_3$ algebra which was found as the asymptotic symmetry algebra at null infinity of three dimensional flat supergravity in 
\cite{Barnich:2014cwa}. Hence, modulo the aforementioned difficulties with the representation theory, the results of this section are expected to apply to minimally supersymmetric BMS$_3$ theories as well, such as the free theory obtained in \cite{Barnich:2015sca}. Holographic realizations \`a la \cite{Hijano:2018nhq,Hijano:2017eii} could be realized in the supergravity model of \cite{Barnich:2014cwa}.

\subsection{The algebra and representations}

The minimal supersymmetric extension of the GCA algebra is given in terms of the bosonic generators $L_n$ and $M_n$ and a single fermionic generator $Q_r$:
\begin{subequations}
\label{SGCA_N=1}
\begin{align}
[L_n, L_m] & = (n-m)L_{m+n} + \frac{c_L}{12} n (n^2-1) \delta_{m+n,0}\,, \\
[L_n, M_m] & = (n-m)M_{m+n} + \frac{c_M}{12}n (n^2-1) \delta_{m+n,0}\,, \\
[L_n,Q_{r}] & = (\tfrac{n}{2}-r)Q_{r+n}\,, \\
\{Q_{r}, Q_{s} \} & = M_{r+s} + \frac{c_M}{6} (r^2-\tfrac14)  \delta_{r+s,0}\,, \\
[M_n, M_m] & = 0 = [M_n, Q_{r}] \,.
\end{align}
\end{subequations}
In the Neveu-Schwarz (NS) sector the label $r$ takes half integer values. We will work exclusively in this sector, because the global subalgebra in the Ramond sector has no non-trivial contribution from the supersymmetry generators. It is generated by $L_{n}$, $M_{n}$ with $n= -1,0,+1$ and $Q_{r}$ with $r= \pm \frac12$.

Another advantage of considering the NS sector is that the primary states are still only labeled by their $M_0$ and $L_0$ weights, hence we can keep calling these states $\ket{\Delta,\xi}$. They have the same properties as in the previous section, except now they are also annihilated by the fermionic lowering operators $Q_r$ with $r\geq 1/2$ and the vacuum state $\ket{0}$ is additionally annihilated by $Q_{-1/2}$. 

The super GCA primary field $\phi_p$ can be combined with its fermionic superpartner $\psi_p(t,x)$ into a superfield $\Phi_p(t,x,\theta)$ depending on a Grassmann-valued coordinate $\theta$
\begin{equation}
\label{eq:decomp}
\Phi_p(t,x,\theta) = \phi_p(t,x) +  \theta \psi_p(t,x) \,.
\end{equation}
The transformation properties of this SGCA primary field are given by
\begin{subequations}
\label{SGCAfieldcom}
\begin{align}
\delta_{L_n}\Phi_p(t,x,\theta) = [L_n, \Phi_p(t,x,\theta)] & = \big[t^{n+1} \partial_t + (n+1)x t^n \partial_x + \frac{n+1}{2} t^n \theta \partial_{\theta} \\ & \qquad   + \xi_p n(n+1)x t^{n-1} + \Delta_p (n+1)t^n \big] \Phi_p(t,x,\theta)\,, \nonumber \\
\delta_{M_n}\Phi_p(t,x,\theta) = [M_n, \Phi_p(t,x,\theta)] & = \big[ t^{n+1}\partial_x + \xi_p (n+1) t^n \big] \Phi_p(t,x,\theta)\,, \\
\delta_{Q_r} \Phi_p(t,x,\theta) = [Q_r, \Phi_p(t,x,\theta)] & = \big[ t^{r+\frac12} \left(\partial_{\theta} - \tfrac12 \theta\partial_x\right) - \xi_p(r+\tfrac12)t^{r-\frac12}\theta\big]\Phi_p(t,x,\theta)\,.
\end{align}
\end{subequations}
This defines the differential operators $\cD_{L_n}, \cD_{M_n}$ and $\cD_{Q_r}$. The definition of the in and out states expressed in terms of the superfield is \begin{equation}\label{N=1inout}
\ket{\Delta_p,\xi_p} = \phi_p(0,0) \ket{0}  = \Phi_p(0,0,0)\ket{0}\,, \qquad \bra{\Delta_p,\xi_p} = \lim_{t \to \infty} t^{2\Delta_p} \bra{0}\Phi(t,0,0)\,. 
\end{equation}
From acting with \eqref{SGCAfieldcom} on the superfield with arbitrary fermionic dependence we can derive the supersymmetry transformations of the primaries:
\begin{equation}
\label{eq:decomp1}
Q_{-1/2}|\Delta_p,\xi_p \rangle = \psi_p(0,0)\ket{0} \equiv |\psi_p \rangle \,, \qquad \;\;  Q_{-1/2} |\psi_p\rangle = \frac12 M_{-1} \ket{\Delta_p,\xi_p}\,,
\end{equation} 
and the fermionic state $\ket{\psi_p}$ has $(L_0,M_0)$ weights $(\Delta_p + \frac12, \xi_p)$.

The SGCA modules now includes descendants obtained by acting with the fermionic raising operators $Q_r$ with $r\leq - 1/2$. The inner products of states for the first three non-trivial levels are
\begin{subequations}
\begin{align}
\label{eq:3levelinnprodN1}
\mathfrak{M}_{1/2} & = \bra{\Delta,\xi}Q_{1/2}Q_{-1/2}\ket{\Delta,\xi} = \xi, \\
\mathfrak{M}_{1}   & = \left( \begin{array}{cc} 2 \Delta & 2\xi \\ 2\xi & 0 \end{array}\right) \,, \\
\mathfrak{M}_{3/2} & = \left[ \begin{array}{c} \bra{\Delta,\xi} Q_{+1/2} L_{+1} \\ \bra{\Delta,\xi} Q_{+3/2} \\ \bra{\Delta,\xi} Q_{+1/2} M_{+1} \end{array} \right] \bigg[ L_{-1}Q_{-1/2}\ket{\Delta,\xi} \; Q_{-3/2}\ket{\Delta,\xi} \; M_{-1}Q_{-1/2} \ket{\Delta,\xi} \bigg]  \nonumber  \\
& = \left( \begin{array}{ccc} (
2 \Delta + 1 )\xi	& 2\xi			& 2 \xi^2	\\ 
2\xi 			& \frac{c_M}{3}+\xi	&	0	\\
2\xi^2			& 0			& 0		
\end{array}\right) \,.
\end{align}
\end{subequations}

\subsection{Correlation functions}
The correlation function of the SGC primaries are obtained straightforwardly from the bosonic ones \eqref{twopt}, \eqref{threept} and \eqref{fourpt} by replacing the difference of the two bosonic coordinates $x_{kl}$ with a supersymmetric generalization
\begin{equation}
\mathbf{x}_{kl} = x_k -x_l - \frac12 \theta_k \theta_l\,.
\end{equation}
This implies that the invariant cross ratio $X$ will now contain fermionic coordinates (and we will denote it as $\mathbf{X}$), while $T$ remains unchanged with respect to \eqref{GCAcrossratios}.
Additionally, there is now the possibility of combining three points into a Grassmann valued invariant 
\begin{equation}\label{Theta}
\Theta_{ijk} = \frac{t_{ij}\theta_k + t_{jk}\theta_i + t_{ki} \theta_j}{\sqrt{t_{ij} t_{jk} t_{ki}}}\,.
\end{equation}
Like the cross ratios, this combination is invariant under the global subalgebra of SGCA. This has a consequence for the four point function, since it is now possible to create a nilpotent bilinear combination of the fermionic cross ratios. Explicitly expanding in this fermionic invariant, the general four point function of four SGCA primaries involves two arbitrary functions of $T$ and $\mathbf{X}$
\begin{align}\label{Superfourpt}
& \langle \Phi_i(t_i,x_i,\theta_i) \Phi_j(t_j,x_j,\theta_j) \Phi_m(t_m,x_m,\theta_m) \Phi_n(t_n,x_n,\theta_n) \rangle  = t_{ij}^{-2\Delta_{i}} t_{jm}^{\Delta_i -\Delta_j - \Delta_m + \Delta_n} \nonumber \\
& t_{jn}^{\Delta_i - \Delta_j + \Delta_m - \Delta_n} t_{mn}^{-\Delta_{i} + \Delta_j - \Delta_m - \Delta_n} e^{-\frac{2\mathbf{x}_{ij} \xi_{i}}{t_{ij}}+\frac{\mathbf{x}_{jm}(\xi_i -\xi_j - \xi_m +\xi_n)}{t_{jm}} + \frac{\mathbf{x}_{jn}(\xi_i -\xi_j + \xi_m -\xi_n)}{t_{jn}} + \frac{\mathbf{x}_{mn} (-\xi_i+\xi_j - \xi_m - \xi_n)}{t_{mn}}} \nonumber \\
& \times \left( \tilde{F}_{0}(T,\mathbf{X}) + \sqrt{1-T} \Theta_{imj}\Theta_{inj} \tilde{F}_{\theta}(T,\mathbf{X}) \right)\,,
\end{align}
where the factor of $\sqrt{1-T}$ is introduced for later convenience. 

Including the supersymmetry generators and the fermionic coordinates, we can use the global subgroup to fix our four pairs of supercoordinates as
\begin{equation}\label{Superpoints}
\{ (t_k,x_k,\theta_k)\} = \{(\infty,0,0), (1,0,0), (t,x,\theta), (0,0,\eta) \}\,,
\end{equation}
where $k = \{i,j,m,n\}$. At these points we have
\begin{equation}
T \to t\,, \qquad  \mathbf{X} \to \mathbf{x} = x - \frac12 \theta \eta\,, \qquad \sqrt{1-T} \Theta_{imj}\Theta_{inj} \to \theta\eta\,.
\end{equation}
The four point function at equal external weights is now simply
\begin{align}\label{SGCAblock}
& \bra{\Delta,\xi}\Phi(1,0,0) \Phi(t,x,\theta) \Phi(0,0,\eta) \ket{0}  = t^{-2\Delta} e^{-2\frac{\mathbf{x}}{t}\xi}(F_0(t,x) + \theta \eta F_{\theta}(t,x))\,.
\end{align}
The function $F_0(t,x)$ and $F_\theta(t,x)$ are in principle arbitrary, but should satisfy certain constraints imposed by crossing symmetry. They are related to the supersymmetric Galilean conformal blocks which we will construct in the next section. In section \ref{se:3.4} we will find the block in the large $c$ expansion.

\subsection{The $\cN=1$ SGC blocks}
\label{subsec:N1cb}
The supersymmetric Galilean conformal blocks (SGC blocks) $\cF^{ji}_{mn}(p|t,x,\theta,\eta)$ are defined by writing the four point function as a sum over primaries:
\begin{align}\label{BlockN1}
G^{ji}_{mn}(t,x,\theta,\eta)&  = \bra{\Delta_i ,\xi_i} \Phi_j(1,0,0) \Phi_m(t,x,\theta) \Phi_n(0,0,\eta) \ket{0} \nonumber \\
& = \sum_p c^p_{ji}c^p_{mn} \cF^{ji}_{mn} (p|t,x,\theta, \eta)\,.
\end{align}
They can be computed analogously to the Galilean conformal blocks derived in section \ref{se:2.2}, by inserting the complete basis of states \eqref{basis} into the four point function \eqref{SGCAblock}. Now, however, we should take into account that $\{N\}$ also contains descendants generated with $Q_{r}$ for $r\leq -1/2$. Whenever we need an explicit basis for $\{N\}$ we will denote it as $\{k\},\{l\},\{r\}$ so that it is implied that
\begin{align}
\Phi^{\{N\}}(0,0,0)\ket{0} & = \ket{\Delta,\xi,\{N\} } = L_{-\{k\}} M_{-\{l\}} Q_{-\{r\}} \ket{\Delta,\xi} \nonumber \\
& = L_{-k_1}\ldots L_{-k_n} M_{-l_1}\ldots M_{-l_m} Q_{-r_1}\ldots Q_{-r_a} \ket{\Delta,\xi}\,.
\end{align}
Like before, the $L$ and $M$ descendants are ordered such that $k_{i} \geq k_{i+1}$ and $l_i \geq l_{i+1}$ but the fermionic descendants satisfy the strict inequality $r_i > r_{i+1}$. This is to avoid an over complete basis, since $Q_{r} Q_{r} = \frac12 M_{2r}$.

The result reads:
\begin{align}\label{N=1block}
\cF^{ji}_{mn} (p|t,x,\theta, \eta) &  = \sum_{\{N\},\{N'\}} \frac{\bra{\Delta_i ,\xi_i} \Phi_j(1,0,0)\ket{\Delta_p,\xi_p,\{N\}}}{c_{ijp}} \mathfrak{M}^{\{N\},\{N'\}} \\
& \qquad \qquad \qquad \qquad \times \frac{\bra{\Delta_p, \xi_p, \{N'\}} \Phi_m(t,x,\theta) \Phi_n(0,0,\eta) \ket{0}}{c_{pmn}} \nonumber \\
& = \sum_{\{N\}, \; |N| \in \mathbb{Z}} \beta_{ji}^{p,\{N\},0} (c^p_{mn})^{-1} \bra{\Delta_p,\xi_p,\{N\}} \Phi_m(t,x,\theta) \Phi_n(0,0,\eta)\ket{0}\,. \label{SGCblock}
\end{align}
Here the second line follows from equation \eqref{beta0s}, where the $\beta$'s are defined in appendix \ref{se:C.2}. Note that with our choice of the coordinates \eqref{Superpoints} the sum above only runs over integer level descendants. The half integer level contributions to the sum drops out because they are identically zero in the first three point function and the (inverse) Kac matrix is block diagonal. The fermionic contribution to the block are contained in the second three point function, due to the fermionic dependence in the primary $\Phi_n(0,0,\eta)\ket{0} = \ket{\Delta_n,\xi_n} + \eta Q_{-1/2}\ket{\Delta_n,\xi_n}$. This complicates the computation of this three point function by the OPE (see appendix \ref{se:C.2} for details). The simplest way to compute this second three point function is in terms of differential operators acting on the three point function of primaries:
\begin{align}
& \bra{\Delta_p,\xi_p,\{N\}} \Phi_m(t,x,\theta) \Phi(0,0,\eta) \ket{0} \nonumber \\ 
& \qquad = \bra{\Delta_p,\xi_p} [Q_{r_a}, \ldots [Q_{r_1}, [M_{l_j} ,\ldots [ M_{l_1} ,[L_{k_i}, \ldots, [L_{k_1}, \Phi_m(t,x,\theta) \Phi_n(0,0,\eta)]]]]]] \ket{0} \nonumber \\ 
& \qquad = \sum_{k=m,n} (-)^{\frac{a}{2}(a-1)}\cD_{L_{k_1}} \ldots \cD_{L_{k_i}} \cD_{M_{l_1}} \ldots \cD_{M_{l_j}} \cD^k_{Q_{r_1}} \ldots \cD^k_{Q_{r_a}}\Big( e^{-\xi_{mnp}\frac{\sx}{t}}t^{-\Delta_{mnp}}\Big)\,,
\end{align}
where in the last line the sum indicate that the differential operator $\cD_{Q_{r_i}}$ should act on both $\Phi_m$ and $\Phi_n$ due to the non-trivial $\eta$ dependence in the last primary.
The sign here appears as we have to exchange the order of $a$ fermionic operators in converting the commutators into differential operators. 
At the end of the day, all fermionic dependence is captured by the supersymmetric coordinate $\sx$ in the exponent. Aside from this exponent up to level 1 the expansion is equivalent to \eqref{blockexp}
\begin{align}\label{Sblockexp}
\cF^{ji}_{mn}(p|t,x,\theta,\eta) =  e^{-\xi_{mnp}\frac{\sx}{t}}t^{-\Delta_{mnp}} & \bigg[1 + \frac{t}{2\xi_p} \left( \Delta_{pji}\xi_{pmn} + \xi_{pji}\Delta_{pmn} - \frac{\xi_{pji}\xi_{pmn} \Delta_p}{\xi_p} \right) \\
& \nonumber + x\, \frac{ \xi_{pji}\xi_{pmn}}{2\xi_p} + \ldots \bigg]\,.
\end{align}

The four point function is built from superfields which are bosonic and hence the order in which we combine primaries with the OPE inside the four point function should be irrelevant. Above we computed it in the channel $ji \leftrightarrow mn$, but we might as well have considered the $ni \leftrightarrow mj$ channel. This crossing symmetry takes the invariant cross ratios to
\begin{equation}
T \to \tilde{T} = \frac{t_{in}t_{mj}}{t_{im}t_{nj}}, \qquad \frac{\mathbf{X}}{T} \to \frac{\tilde{\mathbf{X}}}{\tilde{T}} = \frac{\mathbf{x}_{in}}{t_{in}} + \frac{\mathbf{x}_{mj}}{t_{mj}} - \frac{\mathbf{x}_{im}}{t_{im}} - \frac{\mathbf{x}_{nj}}{t_{n}} \,.
\end{equation}
Under crossing symmetry the global SGCA invariant $ \sqrt{1-T}\Theta_{imj} \Theta_{inj}$ in the four point function \eqref{Superfourpt} becomes
\begin{equation}
\Theta_{imj} \to \Theta_{imn}\,, \qquad \Theta_{inj} \to \Theta_{ijn}\,, \qquad 
 \sqrt{1-T}\Theta_{imj} \Theta_{inj} \to \sqrt{T} \Theta_{imn} \Theta_{ijn}\,.
\end{equation}
At the special value of the coordinates \eqref{Superpoints} these new cross ratios become:
\begin{align}
& \tilde{T} = 1-t\,, \qquad \tilde{\mathbf{X}} =  -x\,, \\
& \Theta_{imn} = \frac{i}{\sqrt{t}}(\theta -\eta)\,, \qquad \Theta_{ijn} = - i \eta\,, \qquad 
\sqrt{T} \Theta_{imn} \Theta_{ijn} = \theta\, \eta \,,
\end{align}
and so the product of the fermionic coordinates stays invariant. Finally, crossing symmetry implies for the SGC blocks that 
\begin{equation}
\sum_p c^p_{ji}c^p_{mn} \cF^{ji}_{mn} (p|t, \mathbf{x}, \theta\, \eta) = \sum_{p'} c^{p'}_{ji} c^{p'}_{mn} \cF^{ni}_{mn}(p'|1-t, -x, \theta\, \eta)\,.
\end{equation}
This is the supersymmetric Galilean conformal bootstrap equation.

\subsection{Global SGC blocks}\label{se:3.4}
Like in the GCA case discussed in the previous section, we would like to find a closed form expression for the SGCA blocks. This is possible in the limit of large central charge by solving the differential Casimir eigenequation on the four point function 
\eqref{SGCAblock}. Let us denote the two arbitrary function $F_{0}(t,x)$ and $F_{\theta}(t,x)$ in the large $c$ limit as
\begin{equation}
F_I (t,x) = \sum_p c^p_{ji}c^p_{mn} f_I(p|x,t) + \cO \left(\frac{\xi_p}{c_M}, \frac{1}{c_L} \right) \,,
\end{equation}
for $I = 0,\theta$.
The differential eigenvalue equations can be obtained by acting with the quadratic Casimirs inside the four point function \eqref{Superfourpt} and taking the coordinates to their special values given in \eqref{Superpoints} afterwards. Now, however, we have more generators in the global subalgebra, $Q_{\pm1/2}$. These generators commute with the first Casimir in equation \eqref{GCACasimirs} and hence $\cC_1$ and its eigenvalue remains unchanged. The equations are hence identical to the bosonic case 
\begin{equation}\label{C1eqn}
\cD_x^2 f_I(p|t,x) = 0\,.
\end{equation}
The second Casimir gets contributions from the supersymmetry generators and is now
\begin{align}
\cC_2  = & \; 2 L_0 M_0 - \frac12 \left( M_{-1}L_{+1} + M_{+1}L_{-1} + L_{-1}M_{+1} + L_{+1}M_{-1} \right) \\
& \nonumber + \frac12 \left(Q_{+1/2} Q_{-1/2} - Q_{-1/2} Q_{+1/2} \right)\,. 
\end{align}
Due to the fermionic contribution, the eigenvalue on SGCA primary states $\ket{\Delta_p,\xi_p}$ is now shifted with respect to the bosonic case. It is
\begin{equation}
\lambda^p_2 = \xi_p (2\Delta_p - \tfrac32)\,.
\end{equation}
Acting inside the four point function \eqref{SGCAblock}, the Casimir $\cC_2$ leads to the following differential equations for the functions $f_I(p|t,x)$:
\begin{subequations}
\begin{align}
\cD_{tx} f_0(p|t,x) & = t f_\theta(p|t,x) \,, \\ 
\cD_{tx} f_\theta(p|t,x) & = \frac{\xi_p^2}{4t}f_0 (p|t,x) - 2 (1-t)t \partial_x f_\theta(p|t,x)\,.
\end{align}
\end{subequations}
with the differential operator $\cD_{tx}$ defined in equation \eqref{DCas}. There are 4 linearly independent solutions to this system of coupled second order differential equations. We are not interested in all of these solutions, merely in the ones which agree with \eqref{Sblockexp} when expanded for small $t$ and $x$. By comparing this with \eqref{Sblockexp} we can fix all four integration constants and obtain
\begin{align}
f_\theta(p|t,x) & = - \frac{\xi_p}{2t}f_0(p|t,x)\,, \\
f_0(p|t,x) & =   \frac{e^{ \xi_p \frac{x}{t\sqrt{1-t}} } }{ \sqrt{1-t} } t^{  \Delta_p } \left( \tfrac12 + \tfrac12\sqrt{1-t} \right)^{2-2\Delta_p}\,.
\end{align}
Using this result, the global $\cN=1$ SGC block with equal external weights $\Delta_i = \Delta$ and $\xi_i = \xi$, defined as
\begin{equation}
g_{\Delta,\xi}(p|t,x,\theta,\eta) = \lim_{\substack{c_L\to \infty \\ c_M/\xi_p \to \infty}} \cF^{ii}_{ii}(p|t,x,\theta,\eta)\,,
\end{equation}
reads
\begin{equation}
g_{\Delta,\xi}(p|t,x,\theta,\eta) =  \frac{e^{ - (2 \xi - \xi_p) \frac{\mathbf{x}}{t} + \xi_p \frac{x}{t}\frac{1-\sqrt{1-t}}{\sqrt{1-t} } }}{t^{ 2 \Delta - \Delta_p } \sqrt{1-t} }  \left( \frac12 + \frac12 \sqrt{1-t} \right)^{2-2\Delta_p} \,.
\end{equation}
Note that all dependence on the fermionic coordinates is in the $\sx$ term in the exponent, but not all $x$ dependence appears as a supersymmetric $\sx$. Aside from this the expression is equivalent to the bosonic block \eqref{bosblock}.

\section{Supersymmetric Galilean conformal blocks: $\cN=2$ democratic}\label{se:4}

In this section we will repeat the procedure for the $\cN=2$ democratic SGCA \eqref{SGCAdemocratic1}, which we repeat here for convenience
\begin{subequations}
\label{SGCAdemocratic}
\begin{align}
[L_n, L_m] & = (n-m)L_{m+n} + \frac{c_L}{12} n (n^2-1) \delta_{m+n,0}\,, \\
[L_n, M_m] & = (n-m)M_{m+n} + \frac{c_M}{12}n (n^2-1) \delta_{m+n,0}\,, \\
[L_n,Q^{\pm}_{r}] & = (\tfrac{n}{2}-r)Q^{\pm}_{r+n}\,, \\
\{Q^{\pm}_{r}, Q^{\pm}_{s} \} & = M_{r+s} + \frac{c_M}{6} (r^2-\tfrac14)  \delta_{r+s,0}\,, \\
\{Q^{\pm}_{r},Q^{\mp}_{s} \} &  = 0 = [M_n, M_m] = [M_n, Q^{\pm}_{r}] \,.
\end{align}
\end{subequations}
We will still restrict ourselves to the NS sector for the reasons outlined in the last section. We also do not consider $R$-symmetry, as we obtained this algebra from a limit of $\cN=(1,1)$ without any $R$-symmetry. This democratic algebra is a simple extension of the $\cN=1$ algebra \eqref{SGCA_N=1}, with doubled fermionic generators $Q_r^\pm$ and superspace Grassmann coordinates $\theta^{\pm}$. The action of the generators on a primary superfield is immediately obtained from the previous section
\begin{subequations}
\label{demSGCAfieldcom}
\begin{align}
\delta_{L_n}\Phi_p(t,x,\theta^\pm) = [L_n, \Phi_p] & = \big[t^{n+1} \partial_t + (n+1)x t^n \partial_x + \frac{n+1}{2} t^n (\theta^+ \partial_{\theta^+} +\theta^- \partial_{\theta^-}) \\ & \qquad   + \xi_p n(n+1)x t^{n-1} + \Delta_p (n+1)t^n \big] \Phi_p(t,x,\theta^\pm)\,, \nonumber \\
\delta_{M_n}\Phi_p(t,x,\theta^\pm) = [M_n, \Phi_p] & = \big[ t^{n+1}\partial_x + \xi_p (n+1) t^n \big] \Phi_p(t,x,\theta^\pm)\,, \\
\delta_{Q^\pm_r} \Phi_p(t,x,\theta^\pm) = [Q^\pm_r, \Phi_p] & = \big[ t^{r+\frac12} \left(\partial_{\theta^\pm} - \tfrac12 \theta^\pm\partial_x\right) - \xi_p(r+\tfrac12)t^{r-\frac12}\theta^\pm\big]\Phi_p(t,x,\theta^\pm)\,.
\end{align}
\end{subequations}
The superfield $\Phi_p$ now contains two fermionic and two bosonic fields
\begin{equation}
\Phi_p(t,x,\theta^{\pm}) = \phi_p(t,x) + \theta^+ \psi_p^+(t,x) + \theta^- \psi_p^-(t,x) + \theta^+ \theta^- F_p(t,x)\,.
\end{equation}
In terms of this superfield, the in and out states are defined with vanishing fermionic coordinates dependence
\begin{equation}
\ket{\Delta_p,\xi_p} = \Phi_p(0,0,0,0)\ket{0}\,, \qquad \bra{\Delta_p,\xi_p} = \lim_{t \to \infty} t^{2\Delta_p} \bra{0}\Phi_p(t,0,0,0)\,. 
\end{equation}
The fermionic states $\ket{\psi^{\pm}} \equiv \psi^{\pm}(0,0)\ket{0} $ and the boson $\ket{F} \equiv F(0,0)\ket{0}$ are supersymmetric descendants of the primary state $\ket{\Delta,\xi}$:
\begin{align}
\ket{\psi^{\pm}} = Q_{-1/2}^{\pm} \ket{\Delta,\xi}\,, \qquad \ket{F} = - Q_{-1/2}^+ Q_{-1/2}^- \ket{\Delta, \xi}\,.
\end{align}
Their $L_0$ weight is shifted by $\frac12$ for the fermions and by one for $\ket{F}$. Since $M_n$ commutes with the supersymmetry generators, their $M_0$ weights are unchanged. The democratic SGCA module now contains raising operators for both fermionic generators
\begin{align}
\ket{\Delta,\xi,\{N\} } & = L_{-\{k\}} M_{-\{l\}} Q^+_{-\{r\}} Q^-_{-\{s\}}\ket{\Delta,\xi} \\
& = L_{-k_1}\ldots L_{-k_n} M_{-l_1}\ldots M_{-l_m} Q^+_{-r_1}\ldots Q^+_{-r_a} Q^-_{-s_1}\ldots Q^-_{-s_b} \ket{\Delta,\xi}\,,
\end{align}
with $s_i > s_{i+1}$. The inner product of states $\mathfrak{M}_{\{N\},\{N'\}} = \bracket{\Delta,\xi,\{N\}}{\Delta,\xi,\{N'\}} $ for the first few levels gives:
\begin{subequations}
\begin{align}
\mathfrak{M}_{1/2} & = \left[ \begin{array}{c}
\bra{\Delta,\xi}Q^+_{1/2} \\
\bra{\Delta,\xi}Q^-_{-1/2}
\end{array} \right]  \left[ Q^+_{-1/2}\ket{\Delta,\xi} \,\,  Q^-_{-1/2}\ket{\Delta,\xi} \right]=  \left( \begin{array}{cc} \xi & 0 \\ 0 & \xi \end{array} \right) \;, \\
\mathfrak{M}_{1} & = \left[ \begin{array}{c}
\bra{\Delta,\xi}L_{1} \\
\bra{\Delta,\xi}M_{1} \\
\bra{\Delta,\xi}Q^-_{-1/2} Q^+_{1/2}
\end{array} \right]  \left[L_{-1}\ket{\Delta,\xi} \,\, M_{-1}\ket{\Delta,\xi} \,\,  Q^+_{-1/2}Q^-_{-1/2}\ket{\Delta,\xi} \right] \nonumber  \\ 
& =  \left( \begin{array}{ccc} 2\Delta & 2\xi & 0 \\ 2\xi & 0 & 0 \\ 0 & 0 & 4 \xi^2 \end{array} \right) \;,
\end{align}
and
\begin{align}
\mathfrak{M}_{3/2}  = \left( \begin{array}{cccccc} 
(2 \Delta + 1 )\xi	& 2\xi			& 2 \xi^2	& 0 & 0 & 0 \\ 
2\xi 			& \frac{c_M}{3}+\xi	&	0	& 0 & 0 & 0\\
2\xi^2			& 0			& 0	& 0 & 0 & 0\\
0 & 0 & 0 & (2 \Delta + 1 )\xi	& 2\xi			& 2 \xi^2	 \\
0 & 0 & 0 & 2\xi 			& \frac{c_M}{3}+\xi	&	0 \\
0 & 0 & 0 & 2\xi^2			& 0			& 0
\end{array}\right) \,,
\end{align}\end{subequations}
in the basis \\ $\{L_{-1}Q^+_{-1/2}\ket{\Delta,\xi},  Q^+_{-3/2}\ket{\Delta,\xi},  M_{-1}Q^+_{-1/2} \ket{\Delta,\xi}, L_{-1}Q^-_{-1/2}\ket{\Delta,\xi}, Q^-_{-3/2}\ket{\Delta,\xi} , M_{-1}Q^-_{-1/2} \ket{\Delta,\xi}  \}$.

\subsection{Correlation functions and an odd sector}
With the algebra extended to two supercharges it becomes possible to construct an invariant of fermionic bilinears in the three point function. This means the general three point function now depends on two structure constants which we will denote as $c_{imn}$ and $\ct_{imn}$
\begin{align}\label{threeptDEM}
& \langle \Phi_i(t_i,x_i,\theta_i^{\pm}) \Phi_m(t_m,x_m,\theta^{\pm}_m) \Phi_n(t_n,x_n,\theta^{\pm}_n) \rangle \nonumber \\ & \qquad \qquad \qquad \qquad \qquad \qquad 
 = \frac{c_{imn} + \Theta^+_{inm}\Theta_{inm}^- \ct_{imn}}{t_{im}^{\Delta_{imn}} t_{mn}^{\Delta_{mni}} t_{ni}^{\Delta_{nim}}}e^{-\frac{\mathbf{x}_{im} \xi_{imn}}{t_{im}}-\frac{\mathbf{x}_{mn} \xi_{mni}}{t_{mn}}-\frac{\mathbf{x}_{ni} \xi_{nim}}{t_{ni}} }\,,
\end{align}
where $\mathbf{x}_{kl} = x_k - x_l - \frac12 \left(\theta^+_k \theta^+_l + \theta^-_k \theta^-_l \right)$ and $\Theta^{\pm}_{inm}$ are given by \eqref{Theta} with $\theta$ replaced by $\theta^\pm$.

At the same time there will be two bosonic and four fermionic cross ratios in the four point function. We take it, at equal external weights, to be
\begin{align}\label{fourptDEM}
& \langle \Phi(t_i,x_i,\theta_i^{\pm}) \Phi(t_j,x_j,\theta_j^{\pm}) \Phi(t_m,x_m,\theta_m^{\pm}) \Phi(t_n,x_n,\theta_n^{\pm}) \rangle  = t_{ij}^{-2\Delta} t_{mn}^{-2\Delta} e^{-2\xi( \mathbf{x}_{ij}/t_{ij} + \mathbf{x}_{mn}/t_{mn})} \nonumber \\
& \times F_{\rm SGCA}(T,\mathbf{X},\Theta^\pm_{imj},\Theta^\pm_{inj})\,.
\end{align}
After a global SGCA transformation, we may fix the coordinates of the four point function to
\begin{equation}\label{SuperpointsDEM}
\{ (t_k,x_k,\theta_k^\pm)
\} = \{(\infty,0,0), (1,0,0), (t,x,\theta^\pm), (0,0,\eta^\pm) \}\,.
\end{equation}
At these points the bosonic cross ratios become $T \to t$ and $\mathbf{X} \to \sx = x - \frac12(\theta^+ \eta^+ + \theta^- \eta^-)$.

Explicitly expanding \eqref{fourptDEM} in the fermionic coordinates, we may write the bosonic function $F_{\rm SGCA}$ in terms of eight arbitrary functions of $t$ and $x$. It is convenient to expand it in terms of the fermionic combinations
\begin{align}
\tau_1^{\pm} & =  \sqrt{1-T}\, \Theta^\pm_{imj} \,, &&
\tau_2^{\pm}  =  \Theta^\pm_{inj}\,,
\end{align}
since at the points \eqref{SuperpointsDEM} these combinations simply become $\tau_1^{\pm} = - \theta^{\pm}$ and $\tau_2^{\pm} = - \eta^\pm$. We will then denote the eight functions appearing in $F_{\rm SGCA}$ as $F_I(t,x)$ where $I$ will denote the Grassmann coordinates they multiply at the points \eqref{SuperpointsDEM}
\begin{align}\label{SGCAblockDEM}
& \bra{\Delta,\xi} \Phi(1,0,0) \Phi(t,x,\theta^{\pm}) \Phi(0,0,\eta^{\pm}) \ket{0} = t^{-2\Delta} e^{-2\frac{\mathbf{x}}{t}\xi} \Big( F_{0}(t,x) + \theta^+ \eta^+ F_{\theta^+ \eta^+}(t,x) \nonumber \\
& \qquad  + \theta^-\eta^+ F_{\theta^-\eta^+}(t,x)  + \theta^-\eta^- F_{\theta^-\eta^-}(t,x) + \theta^+\eta^- F_{\theta^+\eta^-}(t,x) + \theta^+\theta^- F_{\theta^+\theta^- }(t,x) \nonumber \\
& \qquad  + \eta^+\eta^- F_{\eta^+\eta^-}(t,x) + \theta^+ \theta^- \eta^+ \eta^- F_{\theta \theta \eta \eta}(t,x) \Big)\,.
\end{align}

\subsection{Democratic SGC blocks}

Due to the two independent structure constants appearing in the three point function, the four point function does not expand in terms of a single block, but instead it contains four independent blocks, depending on the structure constants they multiply
\begin{align}\label{blocksDEM}
& \bra{\Delta_i,\xi_i} \Phi_j(1,0,0,0) \Phi_m(t,x,\theta^{\pm}) \Phi_n(0,0,\eta^\pm)\ket{0} = \sum_p \Big[ c^p_{ji}c^p_{mn} \cA^{ji}_{mn}(p|t,x,\theta^{\pm},\eta^{\pm}) \\
& \qquad + \ct^p_{ji} c^p_{mn} \cB^{ji}_{mn}(p|t,x,\theta^{\pm},\eta^{\pm}) + c^p_{ji} \ct^p_{mn} \cC^{ji}_{mn}(p|t,x,\theta^{\pm},\eta^{\pm}) + \ct^p_{ji} \ct^p_{mn} \cD^{ji}_{mn}(p|t,x,\theta^{\pm},\eta^{\pm}) \Big]\,. \nonumber
\end{align}
In order to write down explicit expressions for the blocks, it is convenient to split the three point function \eqref{threeptDEM} into two parts, one for each independent structure constant,  or explicitly in terms of conveniently chosen coordinates:
\begin{subequations}
\label{threeptcct}
\begin{align}
\bra{\Delta_i,\xi_i} \Phi_m(t,x,\theta^{\pm}) \Phi_n(0,0,\eta^{\pm}) \ket{0}_c & = \frac{c_{imn}}{t^{\Delta_{mni}}}e^{-\frac{\sx}{t}\xi_{mni}} \,, \\
\bra{\Delta_i,\xi_i} \Phi_m(t,x,\theta^{\pm}) \Phi_n(0,0,\eta^{\pm}) \ket{0}_{\ct} & = \frac{\ct_{imn}}{t^{\Delta_{mni}+1}}e^{-\frac{\sx}{t}\xi_{mni}}  (\theta^+ - \eta^+)(\theta^- - \eta^-)\,.
\end{align} 
\end{subequations}
When inserting the complete basis of states in the four point function \eqref{blocksDEM} and splitting the three point functions as above, we can identify the four separate blocks as
\begin{subequations}
\label{blocks}
\begin{align}\label{Ablock}
\cA^{ji}_{mn}(p|t,x,\theta^{\pm},\eta^{\pm})  &  = \sum_{\{N\},\{N'\}} \frac{\bra{\Delta_i ,\xi_i} \Phi_j(1,0,0)\ket{\Delta_p,\xi_p,\{N\}}_c}{c_{ijp}} \mathfrak{M}^{\{N\},\{N'\}} \\
& \qquad \qquad \qquad \qquad \times \frac{\bra{\Delta_p, \xi_p, \{N'\}} \Phi_m(t,x,\theta^\pm) \Phi_n(0,0,\eta^\pm) \ket{0}_c}{c_{pmn}} \nonumber \\
& \nonumber = \sum_{\{N\}, \; |N| \in \mathbb{Z}} \beta_{ji}^{p,\{N\},0} (c^p_{mn})^{-1} \bra{\Delta_p,\xi_p,\{N\}} \Phi_m(t,x,\theta^\pm) \Phi_n(0,0,\eta^\pm)\ket{0}_c\,, \\
\label{Bblock} \cB^{ji}_{mn}(p|t,x,\theta^{\pm},\eta^{\pm}) &  = \sum_{\{N\},\{N'\}} \frac{\bra{\Delta_i ,\xi_i} \Phi_j(1,0,0)\ket{\Delta_p,\xi_p,\{N\}}_{\ct}}{\ct_{ijp}} \mathfrak{M}^{\{N\},\{N'\}} \\
& \qquad \qquad \qquad \qquad \times \frac{\bra{\Delta_p, \xi_p, \{N'\}} \Phi_m(t,x,\theta^\pm) \Phi_n(0,0,\eta^\pm) \ket{0}_c}{c_{pmn}} \nonumber \\
& \nonumber = \sum_{\{N\}, \; |N| \in \mathbb{Z}}  \bt_{ji}^{p,\{N\},0} (c^p_{mn})^{-1} \bra{\Delta_p,\xi_p,\{N\}} \Phi_m(t,x,\theta^\pm) \Phi_n(0,0,\eta^\pm)\ket{0}_{c}\,, \\
\label{Cblock} \cC^{ji}_{mn}(p|t,x,\theta^{\pm},\eta^{\pm}) & = \sum_{\{N\},\{N'\}} \frac{\bra{\Delta_i ,\xi_i} \Phi_j(1,0,0)\ket{\Delta_p,\xi_p,\{N\}}_c}{c_{ijp}} \mathfrak{M}^{\{N\},\{N'\}} \\
& \qquad \qquad \qquad \qquad \times \frac{\bra{\Delta_p, \xi_p, \{N'\}} \Phi_m(t,x,\theta^\pm) \Phi_n(0,0,\eta^\pm) \ket{0}_{\ct}}{\ct_{pmn}} \nonumber \\
& \nonumber = \sum_{\{N\}, \; |N| \in \mathbb{Z}} \beta_{ji}^{p,\{N\},0} (\ct^p_{mn})^{-1} \bra{\Delta_p,\xi_p,\{N\}} \Phi_m(t,x,\theta^\pm) \Phi_n(0,0,\eta^\pm)\ket{0}_{\ct}\,,
\end{align}
\begin{align}
\label{Dblock} \cD^{ji}_{mn}(p|t,x,\theta^{\pm},\eta^{\pm}) & = \sum_{\{N\},\{N'\}} \frac{\bra{\Delta_i ,\xi_i} \Phi_j(1,0,0)\ket{\Delta_p,\xi_p,\{N\}}_{\ct}}{\ct_{ijp}} \mathfrak{M}^{\{N\},\{N'\}} \\
& \qquad \qquad \qquad \qquad \times \frac{\bra{\Delta_p, \xi_p, \{N'\}} \Phi_m(t,x,\theta^\pm) \Phi_n(0,0,\eta^\pm) \ket{0}_{\ct}}{\ct_{pmn}} \nonumber \\
& = \sum_{\{N\}, \; |N| \in \mathbb{Z}} \bt_{ji}^{p,\{N\},0} (\ct^p_{mn})^{-1} \bra{\Delta_p,\xi_p,\{N\}} \Phi_m(t,x,\theta^\pm) \Phi_n(0,0,\eta^\pm)\ket{0}_{\ct}\,. \nonumber
\end{align}
\end{subequations}
Here the OPE coefficients $\beta$ and $\bt$ are defined in appendix \ref{sec:OPE} and they can be computed recursively. It is shown there that there is no mixing between the $\beta$ and $\bt$ recursive relations and so they define two independent sectors.
In order to compute these blocks explicitly as a series expansion in $t$ and $x$ we can compute the three point functions appearing in \eqref{blocks} by acting with the differential operators \eqref{demSGCAfieldcom} on the appropriate three point functions \eqref{threeptcct}. The first few terms in the expansion of the blocks at equal external weights read
\begin{subequations}
\label{DEMblockexp}
\begin{align}
\cA(p|t,x,\theta^\pm,\eta^\pm)  = t^{-2\Delta + \Delta_p} e^{-\frac{\sx}{t}(2\xi-\xi_p)} & \left[1 + t \frac{\Delta_p}{2} + x \frac{\xi_p}{2} + \ldots \right]\,,  \\
\cB(p|t,x,\theta^\pm,\eta^\pm)  = t^{-2\Delta + \Delta_p} e^{-\frac{x}{t}(2\xi-\xi_p)} & \left[ - \tfrac14 (\theta^+ + \eta^+)(\theta^- + \eta^-) + \ldots  \right]\,, \\
\cC(p|t,x,\theta^\pm,\eta^\pm)  = t^{-2\Delta + \Delta_p-1} e^{-\frac{x}{t}(2\xi-\xi_p)} & \Big[(1+ t \tfrac{\Delta_p}{2} + x \tfrac{\xi_p}{2})(\theta^+ - \eta^+)(\theta^- - \eta^-) \\ & \quad + \tfrac{t}{2} \theta^+\theta^- - \tfrac{t}{2}\eta^+\eta^- + \ldots \Big]\,, \nonumber \\
\cD(p|t,x,\theta^\pm,\eta^\pm)  =  t^{-2\Delta + \Delta_p} e^{-\frac{\sx}{t}(2\xi+\xi_p)} & \frac{t  e^{2\frac{x}{t} \xi_p} }{\xi_p^2} \left[1 + t \frac{1+\Delta_p}{2} + x \frac{\xi_p}{2} + \ldots \right]\,.
\end{align}
\end{subequations}

By considering a different exchange channel for the four point function \eqref{fourptDEM}, for instance $ni \leftrightarrow mj$, we can formulate the $\cN=2$ democratic bootstrap equations. Choosing this different channel takes the invariant cross ratios $T, \mathbf{X}, \tau_1^\pm $ and $\tau_2^\pm$ at the points \eqref{SuperpointsDEM} to
\begin{equation}
T \rightarrow 1-t \,,\qquad \mathbf{X} \rightarrow -x \,, \qquad \tau_1^\pm \rightarrow i (\theta^\pm - \eta^{\pm}) \,, \qquad \tau_2^\pm \rightarrow - i \eta^\pm\,,
\end{equation}
and so the four point function \eqref{blocksDEM} can equally well be expanded as
\begin{align}
& \sum_{p'}  \Big[ c^{p'}_{ni}c^{p'}_{mj} \cA^{ni}_{mj}(p'|1-t,-x,i(\eta^{\pm} - \theta^{\pm}),i\eta^{\pm}) + \ct^{p'}_{ni} c^{p'}_{mj} \cB^{ni}_{mj}(p'|1-t,-x,i(\eta^{\pm} - \theta^{\pm}),i\eta^{\pm}) \nonumber \\
& + c^{p'}_{ni} \ct^{p'}_{mj} \cC^{ni}_{mj}(p'|1-t,-x,i(\eta^{\pm} - \theta^{\pm}),i\eta^{\pm})  + \ct^{p'}_{ni} \ct^{p'}_{mj} \cD^{ni}_{mj}(p'|1-t,-x,i(\eta^{\pm} - \theta^{\pm}),i\eta^{\pm})  \Big]  \,.
\end{align}
Comparing like powers of the independent structure constants $c$ and $\ct$ in this expression and \eqref{blocksDEM} allows us to formulate the expressions
\begin{align}
\sum_p c^p_{ji}c^p_{mn} \cA^{ji}_{mn}(p|t,\sx,\theta^{\pm},\eta^{\pm}) & =  \sum_{p'} c^{p'}_{ni}c^{p'}_{mj} \cA^{ni}_{mj}(p'|1-t,-x,i(\eta^{\pm} - \theta^{\pm}),i\eta^{\pm})\,, \\
\sum_p \ct^p_{ji} \ct^p_{mn} \cD^{ji}_{mn}(p|t,\sx,\theta^{\pm},\eta^{\pm}) & =  \sum_{p'} \ct^{p'}_{ni} \ct^{p'}_{mj} \cD^{ni}_{mj}(p'|1-t,-x,i(\eta^{\pm} - \theta^{\pm}),i\eta^{\pm}) \,,
\end{align}
and
\begin{align}
&\sum_p  \Big[ \ct^p_{ji} c^p_{mn} \cB^{ji}_{mn}(p|t,\sx,\theta^{\pm},\eta^{\pm}) + c^p_{ji} \ct^p_{mn} \cC^{ji}_{mn}(p|t,\sx,\theta^{\pm},\eta^{\pm}) \Big]  \\
& =  \sum_{p'} \Big[ \ct^{p'}_{ni} c^{p'}_{mj} \cB^{ni}_{mj}(p'|1-t,-x,i(\eta^{\pm} - \theta^{\pm}),i\eta^{\pm}) + c^{p'}_{ni} \ct^{p'}_{mj} \cC^{ni}_{mj}(p'|1-t,-x,i(\eta^{\pm} - \theta^{\pm}),i\eta^{\pm}) \Big] \nonumber\,.
\end{align}
These are the bootstrap equations for the democratic (or homogeneous) $\cN=2$ supersymmetric Galilean conformal field theories.

\subsection{Global $\cN =2$ democratic blocks}

In the large $c$ limit we can again find closed form expressions for the blocks \eqref{blocksDEM} by acting with the Casimir of the global subalgebra as a differential operator acting on the four point function and solving the corresponding differential eigenvalue equations. Each of the four blocks in \eqref{blocksDEM} will solve the same differential equations, so for now we will not yet distinguish between them, but simply write the large $c$ limit of any of the blocks as $g_{\Delta,\xi}(p|t,x,\theta^\pm,\eta^\pm)$. After finding the most general solution to the differential equations we can compare with the explicit expansion \eqref{DEMblockexp} and fix the integration constants for each of the blocks. 

We now proceed to find the differential equations for the blocks. First we expand the function $g_{\Delta,\xi}(p|t,x,\theta^\pm,\eta^\pm)$ in the fermionic coordinates in the same way as in \eqref{SGCAblockDEM}
\begin{align}\label{SGCAgbDEM}
& g_{\Delta,\xi}(p|t,x,\theta^{\pm},\eta^{\pm}) = t^{-2\Delta} e^{-2\frac{\mathbf{x}}{t}\xi} \Big( f_{0}(p|t,x) + \theta^+ \eta^+ f_{\theta^+ \eta^+}(p|t,x) + \theta^-\eta^+ f_{\theta^-\eta^+}(p|t,x)  \\
&  \qquad + \theta^-\eta^- f_{\theta^-\eta^-}(p|t,x) \nonumber + \theta^+\eta^- f_{\theta^+\eta^-}(p|t,x) + \theta^+ \theta^- f_{\theta^+ \theta^-}(p|t,x) \\
&\ \qquad \nonumber + \eta^+\eta^- f_{\eta^+\eta^-}(p|t,x)  + \theta^+ \theta^- \eta^+ \eta^- f_{\theta\theta\eta\eta}(p|t,x) \Big)\,.
\end{align}
The first Casimir is unchanged with respect to $\cC_1$ in \eqref{GCACasimirs} and hence every function $f_I(p|t,x)$ above solves the differential equation
\begin{equation}\label{C1eqnDEM}
\cD_x^2 f_I(p|t,x) = 0\,.
\end{equation}
The second Casimir now gets contributions from both supercharges, which changes its eigenvalue to $\lambda^p_2 = \xi_p(2\Delta_p -1)$. 

Just as the blocks and the OPE, also the Casimir equations decouple into two sectors. The functions $f_0,f_{\theta^+ \eta^+},f_{\theta^-\eta^-}$ and $f_{\theta \theta \eta \eta}$ obey one set of coupled differential equations and the rest (the functions which only multiply the mixed $\pm$ fermionic terms in \eqref{SGCAgbDEM}) obey another set. The first set of equations is
\begin{subequations}
\begin{align}
\cD_{tx} f_0 & = t(f_{\theta^+ \eta^+} +f_{\theta^-\eta^-}) \,, \\
\cD_{tx} f_{\theta^+ \eta^+}  & = (1-t)t\left( \frac14 \partial_x^2 f_0 - 2 \partial_x  f_{\theta^+ \eta^+} \right) - tf_{\theta \theta \eta \eta} \,, \\
\cD_{tx} f_{\theta^-\eta^-}  & = (1-t)t\left( \frac14 \partial_x^2 f_0 - 2 \partial_x  f_{\theta^-\eta^-} \right) - tf_{\theta \theta \eta \eta} \,, \\
\cD_{tx} f_{\theta \theta \eta \eta} & = - \frac14(1-t)t\left( \partial_x^2 (f_{\theta^+ \eta^+} + f_{\theta^-\eta^-}) + 16\partial_x f_{\theta \theta \eta \eta} \right)\,.
\end{align}
\end{subequations}
The solution to this set of equations (and simultaneously to \eqref{C1eqnDEM}) can be parameterized by four arbitrary constants, $a_{1,2,3,4}$ \footnote{Actually we are discarding another four independent solutions from the beginning because they solve \eqref{C1eqnDEM} with the wrong sign in the exponent as compared with the OPE.}
\begin{subequations}
\begin{align}
f_0 & = \frac{e^{\frac{\xi_p x}{t\sqrt{1-t}}}}{\sqrt{1-t}} t^{\Delta_p} (1+\sqrt{1-t})^{2-2\Delta_p} \left[a_1 + a_2 \frac{1-\sqrt{1-t}}{1+\sqrt{1-t}} + \frac{a_3 \sqrt{t}}{1+\sqrt{1-t}} \right]\,, \\
f_{\theta^+ \eta^+} & = -\frac{\xi_p}{2t}\frac{e^{\frac{\xi_p x}{t\sqrt{1-t}}}}{\sqrt{1-t}} t^{\Delta_p} (1+\sqrt{1-t})^{2-2\Delta_p} \left[a_1 - a_2 \frac{1-\sqrt{1-t}}{1+\sqrt{1-t}}  - \frac{a_4 \sqrt{t}}{1+\sqrt{1-t}} \right]\,, \\
f_{\theta^-\eta^-} & = -\frac{\xi_p}{2t}\frac{e^{\frac{\xi_p x}{t\sqrt{1-t}}}}{\sqrt{1-t}} t^{\Delta_p} (1+\sqrt{1-t})^{2-2\Delta_p} \left[a_1 - a_2 \frac{1-\sqrt{1-t}}{1+\sqrt{1-t}}  + \frac{a_4 \sqrt{t}}{1+\sqrt{1-t}} \right]\,, \\
f_{\theta \theta \eta \eta} & = -\frac{\xi_p^2}{4t^2}\frac{e^{\frac{\xi_p x}{t\sqrt{1-t}}}}{\sqrt{1-t}} t^{\Delta_p} (1+\sqrt{1-t})^{2-2\Delta_p} \left[a_1 + a_2\frac{1-\sqrt{1-t}}{1+\sqrt{1-t}} - \frac{a_3 \sqrt{t}}{1+\sqrt{1-t}} \right]\,.
\end{align}
\end{subequations}
The second set of equations is
\begin{subequations}
\begin{align}
\cD_{tx} f_{\theta^-\eta^+} & =
- \frac12 (1-t) t \partial_x f_{\theta^+\theta^- } - \frac12 t \partial_x f_{\eta^+\eta^-} -2 (1-t)t \partial_x  f_{\theta^-\eta^+}\,, \\
\cD_{tx} f_{\theta^+\eta^-} & =
 \frac12 (1-t) t \partial_x f_{\theta^+\theta^- } + \frac12 t \partial_x f_{\eta^+\eta^-} - 2 (1-t)t \partial_x  f_{\theta^+\eta^-}\,, \\
\cD_{tx} f_{\theta^+\theta^- } & = - \frac12 t \partial_x (f_{\theta^-\eta^+} - f_{\theta^+\eta^-}) - (2-3t)t\partial_x f_{\theta^+\theta^- }\,, \\
\cD_{tx} f_{\eta^+\eta^-} & = - \frac12(1-t)t\partial_x (f_{\theta^-\eta^+} - f_{\theta^+\eta^-}) - (2-t)t \partial_x f_{\eta^+\eta^-}\,.
\end{align}
\end{subequations}
The solution to these equations (and simultaneously to \eqref{C1eqnDEM}) are parameterized by the arbitrary constants, $b_{1,2,3,4}$
\begin{subequations}
\begin{align}
f_{\theta^-\eta^+} & = \frac{e^{\frac{\xi_p x}{t\sqrt{1-t}}}}{\sqrt{1-t}} t^{\Delta_p-1} (1+\sqrt{1-t})^{2-2\Delta_p} \left[b_1 - \frac{b_2(1-\sqrt{1-t})}{1+\sqrt{1-t}} + \frac{b_3 \sqrt{t}}{1+\sqrt{1-t}} \right]\,, \\
f_{\theta^+\eta^-} & = \frac{e^{\frac{\xi_p x}{t\sqrt{1-t}}}}{\sqrt{1-t}} t^{\Delta_p-1} (1+\sqrt{1-t})^{2-2\Delta_p} \left[- b_1 + \frac{b_2(1-\sqrt{1-t})}{1+\sqrt{1-t}} + \frac{b_3 \sqrt{t}}{1+\sqrt{1-t}} \right]\,,  \\
f_{\theta^+\theta^- } & = \frac{e^{\frac{\xi_p x}{t\sqrt{1-t}}}}{1-t} t^{\Delta_p-1} (1+\sqrt{1-t})^{2-2\Delta_p} \left[ b_1 + \frac{b_2(1-\sqrt{1-t})}{1+\sqrt{1-t}} + \frac{b_4 \sqrt{t}}{1+\sqrt{1-t}} \right]\,, \\
f_{\eta^+\eta^-} & = e^{\frac{\xi_p x}{t\sqrt{1-t}}} t^{\Delta_p-1} (1+\sqrt{1-t})^{2-2\Delta_p} \left[ b_1 + \frac{b_2(1-\sqrt{1-t})}{1+\sqrt{1-t}} - \frac{b_4 \sqrt{t}}{1+\sqrt{1-t}} \right]\,.
\end{align}
\end{subequations}
From the structure of the fermionic terms (see for instance \eqref{DEMblockexp}) it is clear that the mixed blocks ($\cB_{mn}^{ji}$ and $\cC_{mn}^{ji}$) should correspond to solutions with non-zero $b$ coefficients, while the other blocks have non-zero $a$ coefficients. We can find the value for these integration constants by comparing the small $t$ and $x$ expansion of the solutions above with (the large $c$ limit of) \eqref{DEMblockexp}. We find that the blocks are given by the above function $g_{\Delta,\xi}(p|t,x,\theta^{\pm}, \eta^{\pm})$ with only one non-zero integration constant. These are for the
\begin{subequations}
\begin{align}
\cA - \text{block}:   \; a_1 & = 2^{-2+2\Delta_p}\,,  & \cD - \text{block}:  \; a_2 & = \frac{2^{2\Delta_p}}{\xi_p^2}\,, \\ 
 \cB - \text{block}:  \; b_2 & =  - 2^{2\Delta_p-2}\,,  & \cC - \text{block}:  \; b_1 & = 	2^{2\Delta_p-2}\,.
\end{align}
\end{subequations}
Putting the above results together, the final solution for the global blocks at equal external weights $\Delta$ and $\xi$ reads
\begin{subequations}
\label{FINblocks}
\begin{align}
\cA(p|t,x,\theta^\pm,\eta^\pm) & =  e^{-\frac{\mathbf{x}}{t}(2\xi-\xi_p) } \frac{e^{\frac{(1-\sqrt{1-t})}{\sqrt{1-t}}\frac{x}{t} \xi_p }  }{t^{ 2 \Delta - \Delta_p } \sqrt{1-t} }  \left( \tfrac12 + \tfrac12 \sqrt{1-t} \right)^{2-2\Delta_p}\,, \\
\cB(p|t,x,\theta^\pm,\eta^\pm) & =  \frac{ e^{- \frac{x}{t}(2\xi - \frac{\xi_p}{\sqrt{1-t}} ) } (\theta^+  + \sqrt{1-t} \, \eta^+)( \theta^- + \sqrt{1-t} \, \eta^-) }{4\,t^{ 2 \Delta - \Delta_p } (t-1)  \left( \tfrac12 + \tfrac12\sqrt{1-t} \right)^{2\Delta_p} } \,, \\
\cC(p|t,x,\theta^\pm,\eta^\pm) & = \frac{e^{- \frac{x}{t}(2\xi - \frac{\xi_p}{\sqrt{1-t}} ) } (\theta^+  - \sqrt{1-t} \, \eta^+)( \theta^- - \sqrt{1-t} \, \eta^-) }{t^{ 2 \Delta - \Delta_p +1}  (1-t)  \left( \tfrac12+ \tfrac12 \sqrt{1-t} \right)^{2\Delta_p-2} } \,, \\
\cD(p|t,x,\theta^\pm,\eta^\pm) & =  e^{-\frac{\mathbf{x}}{t}(2\xi+\xi_p) } \frac{e^{\frac{(1+\sqrt{1-t})}{\sqrt{1-t}}\frac{x}{t} \xi_p }  }{\xi_p^2 \, t^{ 2 \Delta - \Delta_p -1 } \sqrt{1-t} }  \left( \tfrac12 + \tfrac12 \sqrt{1-t} \right)^{-2\Delta_p}\,. 
\end{align}
\end{subequations}
This is one of the main results of this paper: the global democratic (or homogeneous) supersymmetric Galilean conformal blocks. Since in this case the global subgroup (which was used to derive this result) is isomorphic to the $\cN=2$ supersymmetric Poincar\'e algebra, these block correspond to the global super Poincar\'e blocks (upon exchanging the coordinates $t $ and $x$). It would be interesting to compute this result holographically in the supergravity theories studied in \cite{Lodato:2016alv} by means of the methods developed for flat space holography in \cite{Hijano:2017eii}.

\section{Supersymmetric Galilean conformal blocks: $\cN=2$ despotic}\label{se:5}

The $\cN=2$ supersymmetric extension of the GCA algebra in which the fermionic generators scale in the same way as the bosonic generators in the limit from the $\cN=(1,1)$ superconformal algebra is referred to as the despotic or the inhomogeneous SGCA algebra. The representations and correlation functions were studied in \cite{Mandal:2010gx}. Here we will extend the analysis to the supersymmetric Galilean blocks and their large $c$ limit.

The algebra has commutation relations
\begin{subequations}
\label{SGCAdespotic}
\begin{align}
[L_n, L_m] & = (n-m)L_{m+n} + \frac{c_L}{12} n (n^2-1) \delta_{m+n,0}\,, \\
[L_n, M_m] & = (n-m)M_{m+n} + \frac{c_M}{12}n (n^2-1) \delta_{m+n,0}\,, \\
[L_n,G_{r}] & = (\tfrac{n}{2}-r)G_{r+n}\,,
\\
 [L_n,H_r]&=[M_n,G_r]= (\tfrac{n}{2}-r)H_{r+n}\,, \\
\{G_{r}, G_{s} \} & = 2\,L_{r+s} + \frac{c_L}{6} (r^2-\tfrac14)  \delta_{r+s,0}\,, \\
\{G_r,H_s\} &= 2\,M_{r+s}+ \frac{c_M}{6} (r^2-\tfrac14)  \delta_{r+s,0}\,,
\\
 [M_n, M_m]&= 0 = [M_n, H_{r}]=\{H_r,H_s\} \,.
\end{align}
\end{subequations}
The algebra has a super Virasoro subalgebra spanned by $L_n$ and $G_r$ and it satisfies a natural grading (apart from the grading of supersymmetry) under which $L_n, G_r$ are even and $M_n, H_r$ are odd. This means that the commutators of even generators among themselves give even generators, even with odd generators give odd generators on the right hand side, while odd generators (anti-)commute. 

We will work exclusively in the NS sector, where $r$ takes half integer values and the global subalgebra is spanned by $L_n, M_n$ with $n=-1,0,+1$ and $G_r, H_r$ with $r= \pm \tfrac12$. The vector fields generating this supersymmetric algebra on superspace parameterized by $(t,x,\theta, \chi)$ are
\begin{subequations}
\begin{align}
L_n&=-t^{n+1}\partial_t-(n+1)x t^n\partial_x-\frac{n+1}{2} t^n \big(\theta\partial_\theta+\chi\partial_\chi\big)-\frac{n(n+1)}{2}x t^{n-1} \theta\partial_\chi \,,
\\
M_n&=-t^{n+1}\partial_x-\frac{n+1}2 t^n \theta\partial_\chi \,, 
\\
G_r&=-t^{r+\frac12}(\theta \partial_t+\chi \partial_x-\partial_\theta)-(r+\tfrac12)x t^{r-\tfrac12}(\theta \partial_x-\partial_\chi) \,, 
\\
H_r&=-t^{r+\frac12}(\theta\partial_x + \partial_\chi)\,,
\end{align}
\end{subequations}
where the Grassmann coordinates $(\theta,\chi)$ are associated to $G_r,H_r$ respectively. Note the appearance of fermionic dependence in the $M_n$ generators, due to their non-trivial commutation relation with the supercharges $G_r$.

Primary fields can be organized as superfields, depending on all super coordinates, which we will keep denoting by $\Phi(t,x,\theta,\chi)$
\begin{equation}\label{despoticfield}
\Phi(t,x,\theta,\chi) = \phi(t,x) +  \theta \psi_{1}(t,x) + \chi \psi_2(t,x)  + \theta \, \chi F(t,x) \,.
\end{equation}
The action of the generators on primary fields is given by
\begin{subequations}
	\label{despSGCAfieldcom}
	\begin{align}
	\delta_{L_n}\Phi_p(t,x,\theta,\chi) & = [L_n, \Phi_p] = \Big[t^{n+1}\partial_t+(n+1)x t^n\partial_x+\frac{n+1}{2} t^n \big(\theta\partial_\theta+\chi\partial_\chi\big) \\ 
	& \;\;  +\frac{n(n+1)}{2}x t^{n-1} \theta\partial_\chi + \xi_p n(n+1)x t^{n-1}
	+ \Delta_p (n+1)t^n \Big] \Phi_p(t,x,\theta,\chi) \,,  \nonumber \\
	\delta_{M_n}\Phi_p(t,x,\theta,\chi) & = [M_n, \Phi_p] = \Big[ t^{n+1}\partial_x+\frac{n+1}2 t^n \theta\partial_\chi+\xi_p (n+1) t^n \Big] \Phi_p(t,x,\theta,\chi)\,, \\
	\delta_{G_r} \Phi_p(t,x,\theta,\chi) & = [G_r, \Phi_p]  = \Big[t^{r+\frac12}(\partial_\theta-\theta \partial_t-\chi \partial_x)+(r+\tfrac12)x t^{r-\frac12}(\partial_\chi-\theta \partial_x)  \\ \nonumber
	& \;\; -2(r+\tfrac12)t^{r-\tfrac12}(\Delta_p \theta+\xi_p \chi) -2 \xi_p (r^2-\tfrac14)t^{r-\frac32} x \theta \Big]\Phi_p(t,x,\theta,\chi) \,,
	\\
	\delta_{H_r}\Phi_p(t,x,\theta,\chi) & = [H_r,\Phi_p ] = \Big[t^{r+\frac12}(\partial_\chi-\theta\partial_x)-2\xi_p (r+\tfrac12) t^{r-\frac12} \theta\Big]\Phi_p(t,x,\theta,\chi)\,.
	\end{align}
\end{subequations}
Primary states are still labeled by their $M_0$ and $L_0$ eigenvalue and the in and out states are defined as before by inserting a field at the origin of superspace and at the infinite future respectively
\begin{equation}\label{despprimary}
\ket{\Delta_p,\xi_p} = \phi_p(0,0)\ket{0} = \Phi_p(0,0,0,0)\ket{0}\,, \qquad \bra{\Delta_p,\xi_p} = \lim_{t \to \infty} t^{2\Delta_p} \bra{0}\Phi_p(t,0,0,0)\,. 
\end{equation}
Here the vacuum state $\ket{0}$ is defined as being annihilated by all lowering operators and the global subgroup (so by all $L_n,M_n,G_n,H_n$ with $n\geq -1$). 
Raising operators $L_n,M_n,G_n,H_n$ with $n < 0$ create descendants in the despotic SGCA module
\begin{align}
\ket{\Delta_p,\xi_p,\{N\}} & \equiv  L_{-\{k\}} M_{-\{l\}} G_{-\{r\}} H_{-\{s\}} \ket{\Delta_p,\xi_p} \\
& = L_{-k_1} \ldots L_{-k_n} M_{-l_1} \ldots M_{-l_m} G_{-r_1} \ldots G_{-r_a} H_{-s_1} \ldots H_{-s_b} \ket{\Delta_p,\xi_p}\,. \nonumber
\end{align} 
Where $\{k\},\{l\},\{r\},\{s\}$ are ordered sets of integers such that $k_i \geq k_{i+1}$, $l_i \geq l_{i+1}$, $r_i > r_{i+1}$ and $s_i > s_{i+1}$. Particularly, following from the transformation rules \eqref{despSGCAfieldcom} the states corresponding to the fields $\psi_1, \psi_2$ and $F$ in \eqref{despoticfield} can be obtained as descendants of \eqref{despprimary}
\begin{subequations}
\begin{align}
\ket{\psi_1} & = G_{-1/2} \ket{\Delta,\xi} \,, \qquad  \ket{\psi_2} = H_{-1/2} \ket{\Delta,\xi} \,,  \\
\ket{F} & = \frac{1}{2}(H_{-1/2} G_{-1/2} - G_{-1/2} H_{-1/2} ) \ket{\Delta, \xi} \,.
\end{align}
\end{subequations}
The $L_0$ weights of these states are shifted by one half for the fermions $\ket{\psi_i}$ and by one for the boson $\ket{F}$, but now $M_0$ acts non-diagonally on the fermions:
\begin{equation}
M_0 \ket{\psi_1} = \xi \ket{\psi_1} + \frac12 \ket{\psi_2} \,, \qquad M_0 \ket{\psi_2} = \xi \ket{\psi_2}\,.
\end{equation}
This is reminiscent of the Jordan block structure encountered in logarithmic CFTs (see for instance \cite{Hogervorst:2016itc}), used in this context in \cite{Hijano:2018nhq} to construct null `BMS-multiplets'.

Hermitian conjugation swaps the order of the generators and takes their index to minus itself.
The inner product of states $\mathfrak{M}_{\{N\},\{N'\}} = \bracket{\Delta,\xi,\{N\}}{\Delta,\xi,\{N'\}} $ for the first few levels gives:
\begin{subequations}
	\begin{align}
	\mathfrak{M}_{1/2} & = \left[ \begin{array}{c}
	\bra{\Delta,\xi}G_{1/2} \\
	\bra{\Delta,\xi}H_{-1/2}
	\end{array} \right]  \left[ G_{-1/2}\ket{\Delta,\xi} \,\,  H_{-1/2}\ket{\Delta,\xi} \right]=  \left( \begin{array}{cc} 2\Delta & 2\xi \\ 2 \xi & 0 \end{array} \right) \;, \\
	\mathfrak{M}_{1} & = \left[ \begin{array}{c}
	\bra{\Delta,\xi}L_{1} \\
	\bra{\Delta,\xi}H_{-1/2} G_{1/2} \\
	\bra{\Delta,\xi}M_{1} 
	\end{array} \right]  \left[L_{-1}\ket{\Delta,\xi}\,\, G_{-1/2} H_{-1/2}\ket{\Delta,\xi}  \,\, M_{-1}\ket{\Delta,\xi} \right] \nonumber  \\ 
	& =  \left( \begin{array}{ccc} 2\Delta & 2\xi & 2\xi \\ 2\xi & 4\xi^2 & 0 \\ 2\xi & 0 & 0 \end{array} \right) \;, 
	\end{align}
and
\begin{align}
\mathfrak{M}_{3/2}  = \left( \begin{array}{cccccc} 
(2 \Delta + 1 )2\Delta	& 4\Delta			& (2 \Delta + 1 )2\xi		& (2 \Delta + 1 )2\xi	 & 4\xi & 4\xi^2 \\ 
4\Delta			& \frac{2}{3}c_L + 2\Delta	&	4\xi	& 	4\xi &  \frac{2}{3}c_M+ 2\xi & 0\\
(2 \Delta + 1 )2\xi  & 	4\xi & 0	&4\xi^2 & 0 & 0\\
(2 \Delta + 1 )2\xi	 & 	4\xi & 4\xi^2 & 0	& 0		& 0	 \\
4\xi    & \frac{2}{3}c_M+ 2\xi	 & 0 &0			& 0 &	0 \\
4\xi^2  & 0 & 0 & 0			& 0			& 0
\end{array}\right) \,,
\end{align}
\end{subequations}
in the basis \\ $\{L_{-1}G_{-1/2}\ket{\Delta,\xi},  G_{-3/2}\ket{\Delta,\xi},  L_{-1}H_{-1/2} \ket{\Delta,\xi}, M_{-1}G_{-1/2}\ket{\Delta,\xi}, H_{-3/2}\ket{\Delta,\xi}, M_{-1}H_{-1/2} \ket{\Delta,\xi}  \}$.

\subsection{Despotic correlation functions and odd sector}
The despotic algebra differs from the democratic algebra only in the fermionic generators, hence the correlation functions for the bosonic quantities remain unchanged, while the fermionic correlators will be different. In the two point function \eqref{twopt} this change is captured entirely by replacing $x_{mn}$ and $t_{mn}$ by the new super coordinates $\sx_{mn}$ and $\st_{mn}$, defined as
\begin{equation}
\sx_{mn} = x_m - x_n -\theta_m \chi_n -\chi_m\theta_n\,, \qquad \text{and} \qquad  {\bf t}_{mn}=t_m-t_n -\theta_m\theta_n\,.
\end{equation}
The generic three point function still depends on two structure constants which we will denote as $c_{imn}$ and $\ct_{imn}$. This time however, the space and time coordinates become the above supersymmetric coordinates and the Grassmann-valued invariants are slightly more complicated. Specifically
\begin{align}\label{threeptDES}
\nonumber & \langle \Phi_i(t_i,x_i,\theta_i,\chi_i) \Phi_m(t_m,x_m,\theta_m,\chi_m) \Phi_n(t_n,x_n,\theta_n,\chi_n) \rangle = \\
& \qquad \qquad \qquad \qquad \qquad\qquad\qquad \frac{c_{imn} + \Theta_{inm}\Xi_{inm} \ct_{imn}}{{\bf t}_{im}^{\Delta_{imn}} {\bf t}_{mn}^{\Delta_{mni}} {\bf t}_{ni}^{\Delta_{nim}}}e^{-\frac{\mathbf{x}_{im} \xi_{imn}}{{\bf t}_{im}}-\frac{\mathbf{x}_{mn} \xi_{mni}}{{\bf t}_{mn}}-\frac{\mathbf{x}_{ni} \xi_{nim}}{{\bf t}_{ni}} }\,.
\end{align}
here $\Theta_{inm}$ is given by \eqref{Theta} with $t$ replaced by $\st$, or explicitly,:
\begin{equation}\label{Thetadesp}
\Theta_{ijk} = \frac{t_{ij}\theta_k + t_{jk}\theta_i + t_{ki} \theta_j - \frac12 \theta_i \theta_j \theta_k}{\sqrt{t_{ij} t_{jk} t_{ki}}}\,,
\end{equation}
and $\Xi_{inm}$ reads:
\begin{align}\label{Xi}
\Xi_{inm}= &(2\sqrt{t_{in} t_{nm} t_{mi}})^{-1}\Big[2(t_{im}\chi_n + t_{mn}\chi_i+t_{ni}\chi_m)+(x_{im}\theta_n +x_{mn}\theta_i+x_{ni}\theta_m) \nonumber
\\
& \qquad -t_{im}\theta_n \Big(\frac{x_{mn}}{t_{mn}}+\frac{x_{ni}}{t_{ni}}\Big)-t_{mn}\theta_i \Big(\frac{x_{ni}}{t_{ni}}+\frac{x_{im}}{t_{im}}\Big)-t_{ni}\theta_m \Big(\frac{x_{im}}{t_{im}}+\frac{x_{mn}}{t_{mn}}\Big)
\nonumber\\
& \qquad -(\theta_i\theta_m\chi_n+\theta_i\chi_m\theta_n+\chi_i\theta_m\theta_n)+\frac12 \theta_i\theta_m\theta_n \Big(\frac{x_{im}}{t_{im}}+\frac{x_{mn}}{t_{mn}}+\frac{x_{ni}}{t_{ni}}\Big)\Big] \,.
\end{align}
It can be verified that these Grassmann valued combinations are invariant under the global subalgebra of the despotic SGCA.\footnote{The possibility to consider these invariants was unfortunately wrongly dismissed in \cite{Mandal:2010gx}.}

The four point correlator depends on an arbitrary function of two bosonic and four fermionic cross ratios. We take it, at equal external weights, to be
\begin{align}\label{fourptDES}
& \langle \Phi(t_i,x_i,\theta_i,\chi_i) \Phi(t_j,x_j,\theta_j,\chi_j) \Phi(t_m,x_m,\theta_m,\chi_m) \Phi(t_n,x_n,\theta_n^{\pm}) \rangle \nonumber \\
& \qquad \qquad \qquad \qquad \qquad = {\bf t}_{ij}^{-2\Delta} {\bf t}_{mn}^{-2\Delta} e^{-2\xi( \mathbf{x}_{ij}/{\bf t}_{ij} + \mathbf{x}_{mn}/{\bf t}_{mn})}  F_{\rm SGCA}(\mathfrak{T},\mathfrak{X},\tau_1,\tau_2,\tau_3,\tau_4)\,.
\end{align}
where now the bosonic cross ratios read
\begin{subequations}
\begin{align}\label{TXDES}
\mathfrak{T}&=\frac{{\bf t}_{ij}{\bf t}_{mn}}{{\bf t}_{im}{\bf t}_{jn}}=\frac{{\bf t}_{ij}{\bf t}_{mn}}{t_{im}t_{jn}} \Big(1-\frac{\theta_j\theta_n}{t_{jn}}-\frac{\theta_i\theta_m}{t_{im}}+\frac{\theta_i\theta_j\theta_m\theta_n}{t_{im}t_{jn}}\Big) \,,
\\
\mathfrak{X}&=\mathfrak{T} \Big[\frac{{\bf x}_{ij}}{{\bf t}_{ij}}+\frac{{\bf x}_{mn}}{{\bf t}_{mn}} -\frac{{\bf x}_{im}}{{\bf t}_{im}}-\frac{{\bf x}_{jn}}{{\bf t}_{jn}} \Big]\,,
\end{align}
\end{subequations}
and we have defined four nilpotent Grassmannian invariants from conveniently chosen combinations of \eqref{Thetadesp} and \eqref{Xi}
\begin{align}
\tau_1 & = \sqrt{1-\mathfrak{T} }\, \Theta_{imj}\,, &&
\tau_2 = \Theta_{inj}\,, \\
\tau_3 & = \sqrt{1-\mathfrak{T} } \left(\Xi_{imj} - \frac12 \frac{\mathfrak{X}}{(1-\mathfrak{T})}  \Theta_{imj} \right)\,, &&	
\tau_4 = \Xi_{inj}\,.
\end{align}
After a global SGCA transformation, we may fix the coordinates of the four point function to
\begin{align}\label{SuperpointsDES}
&\{ (t_k,x_k,\theta_k,\chi_k)\}= \{(\infty,0,0,0), (1,0,0,0), (t,x,\theta,\chi), (0,0,\tilde{\theta},\tilde{\chi}) \}\,.
\end{align}
At these points the cross ratios \eqref{TXDES} become
\begin{equation}
\mathfrak{T} = \st = t - \theta \tilde{\theta} \,, \qquad \qquad \mathfrak{X} = \sx = x - \theta \tilde{\chi} - \chi \tilde{\theta}\,,
\end{equation}
and 
\begin{align}
\tau_1 = - \theta\,, &&
\tau_2 = - \tilde{\theta}\,, &&
\tau_3 = - \chi\,, &&
\tau_4 = - \tilde{\chi}\,.
\end{align}
Explicitly expanding in the fermionic coordinates, we may write the correlator \eqref{fourptDES} in terms of eight bosonic functions. We will label these functions by the Grassmann coordinates they multiply at the points \eqref{SuperpointsDES} as
\begin{align}\label{SGCAblockDES}
& \bra{\Delta,\xi} \Phi(1,0,0) \Phi(t,x,\theta,\chi) \Phi(0,0,\tilde{\theta},\tilde{\chi}) \ket{0} = \st^{-2\Delta} e^{-2\frac{\mathbf{x}}{\st}\xi} \Big( F_{0}(t,x) + \theta \tilde{\theta} F_{\theta \tilde{\theta}}(t,x) + \chi\tilde{\theta} F_{\chi\tilde{\theta}}(t,x) \nonumber \\
& \qquad  + \chi\tilde{\chi} F_{\chi\tilde{\chi}}(t,x) + \theta\tilde{\chi} F_{ \theta\tilde{\chi}}(t,x) + \theta\chi F_{\theta\chi}(t,x) + \tilde{\theta}\tilde{\chi} F_{\tilde{\theta}\tilde{\chi}}(t,x) + \theta \chi\tilde{\theta}\tilde{\chi} F_{\theta \chi\tilde{\theta}\tilde{\chi}}(t,x) \Big)\,, 
\end{align}

\subsection{Despotic SGC blocks}

Just like in the democratic case \eqref{blocksDEM}, the presence of two independent structure constants in the three point function implies that the four point function expands into four blocks
\begin{align}\label{blocksDES}
& \bra{\Delta_i,\xi_i} \Phi_j(1,0,0,0) \Phi_m(t,x,\theta,\chi)\Phi_n(0,0,\tilde{\theta},\tilde{\chi}) \ket{0,0} = \sum_p \Big[ c^p_{ji}c^p_{mn} \cA^{ji}_{mn}(p|\mathbf{Z}) \nonumber \\
& \qquad + \ct^p_{ji} c^p_{mn} \cB^{ji}_{mn}(p|\mathbf{Z}) + c^p_{ji} \ct^p_{mn} \cC^{ji}_{mn}(p|\mathbf{Z}) + \ct^p_{ji} \ct^p_{mn} \cD^{ji}_{mn}(p|\mathbf{Z}) \Big],
\end{align}
where for sake of conciseness we have used $\mathbf{Z}=\{t,x,\theta,\tilde{\theta},\chi,\tilde{\chi}\}$ to denote the coordinate dependence. In order to write down the explicit form of the blocks we split the three point functions \eqref{threeptDES} into the two sectors, one per independent structure constant, i.e. $\langle \Phi\Phi\Phi\rangle = \langle \Phi\Phi\Phi\rangle_c + \langle \Phi\Phi\Phi\rangle_{\ct}$. At a convenient choice of coordinates the three point functions are
\begin{subequations}
\label{threeptcctDES}
	\begin{align}
	\bra{\Delta_i,\xi_i} \Phi_m(t,x,\theta,\chi) \Phi_n(0,0,\thetat,\chit) \ket{0}_c & = \frac{c_{imn}}{\st^{\Delta_{mni}}}e^{-\frac{\sx}{\st}\xi_{mni}} \,, \\
	\bra{\Delta_i,\xi_i} \Phi_m(t,x,\theta,\chi) \Phi_n(0,0,\thetat,\chit) \ket{0}_{\ct} & = \frac{\ct_{imn}}{\st^{\Delta_{mni}+1}}e^{-\frac{\sx}{\st}\xi_{mni}}  (\theta - \thetat)(\chi - \chit)\,.
	\end{align} 
\end{subequations}
The four different blocks are computed the same way as in previous sections: one introduces a complete basis of states in the middle of the four point function \eqref{fourptDES} and identifies the appropriate block by the structure constants it multiplies. The general expressions in this case are equivalent to \eqref{blocks}, but with $\theta^{\pm}$ replaced by $\{\theta,\chi\}$ and $\eta^\pm$ replaced by $\{ \thetat, \chit \}$. Of course, the actual values for the inverse Kac determinant and the OPE coefficients are now different and the three point functions are computed by acting with the differential operators \eqref{despSGCAfieldcom} on the three point functions \eqref{threeptcctDES}. Explicitly computing the first few terms in the expansion gives at equal external weights
\begin{subequations}
\label{DESblockexp}
\begin{align}
	\cA(p|\mathbf{Z}) &  = \st^{-2\Delta + \Delta_p} e^{-\frac{\sx}{\st}(2\xi-\xi_p)}  \left[1 + t \frac{\Delta_p}{2} + x \frac{\xi_p}{2} + \ldots \right]\,,  \\
	\cB(p|\mathbf{Z}) & = \st^{-2\Delta + \Delta_p} e^{-\frac{x}{t}(2\xi-\xi_p)}  \left[ - \tfrac14 (\theta + \thetat)(\chi + \chit) + \ldots  \right]\,, \\
	\cC(p|\mathbf{Z}) & = \st^{-2\Delta - \Delta_p+1} e^{-\frac{\sx}{\st}(2\xi+\xi_p)}  \frac{e^{2\frac{x}{t}\xi_p}}{2} t^{2\Delta_p-2} \Big[  x \theta \thetat + t(\theta \chi - \thetat \chit) \\ & \qquad \qquad \qquad \qquad \qquad \qquad  + (2 + t \Delta_p + x \xi_p)(\theta - \thetat)(\chi - \chit)  + \ldots \Big] \,,\nonumber \\
	\cD(p|\mathbf{Z}) & = - \st^{-2\Delta - \Delta_p+1} e^{-\frac{\sx}{\st}(2\xi+\xi_p)} \frac{t^{2\Delta}  e^{2\frac{x}{t} \xi_p} }{4\xi_p^2} \left[1 + t \frac{1+\Delta_p}{2} + x \frac{\xi_p}{2} + \ldots \right]\,.
\end{align}
\end{subequations}
By considering how the invariant cross ratios behave under crossing symmetry we can obtain the $\cN=2$ despotic bootstrap equations. Crossing symmetry takes the cross ratios $\mathfrak{T}, \mathfrak{X}$ and $ \tau_{k}$ with $k = 1,\ldots,4$ at the points \eqref{SuperpointsDES} to
\begin{subequations}
\begin{align}
&& \mathfrak{T}  & \rightarrow 1-t \,, & \mathfrak{X} & \rightarrow -x \,, \\
\tau_1  \rightarrow i (\theta - \thetat)\,, &&  \tau_2 & \rightarrow - i \thetat \,, & \tau_3 & \rightarrow i(\chi -\chit) \,, && \tau_4 \rightarrow -  i \chit \,.
\end{align}
\end{subequations}
After expanding the four point function \eqref{blocksDES} in the $ni \leftrightarrow mj$ channel and using the above relations we may write the $\cN=2$ despotic (on inhomogeneous) supersymmetric Galilean conformal bootstrap equations as
\begin{align}
\sum_p c^p_{ji}c^p_{mn} \cA^{ji}_{mn}(p|\mathbf{Z}) & =  \sum_{p'} c^{p'}_{ni}c^{p'}_{mj} \cA^{ni}_{mj}(p'|\mathbf{Z'})\,, \\
\sum_p \ct^p_{ji} \ct^p_{mn} \cD^{ji}_{mn}(p|\mathbf{Z}) & =  \sum_{p'} \ct^{p'}_{ni} \ct^{p'}_{mj} \cD^{ni}_{mj}(p'|\mathbf{Z'}) \,,
\end{align}
and
\begin{equation}
\sum_p  \Big[ \ct^p_{ji} c^p_{mn} \cB^{ji}_{mn}(p|\mathbf{Z}) + c^p_{ji} \ct^p_{mn} \cC^{ji}_{mn}(p|\mathbf{Z}) \Big]  =  \sum_{p'} \Big[ \ct^{p'}_{ni} c^{p'}_{mj} \cB^{ni}_{mj}(p'|\mathbf{Z'}) + c^{p'}_{ni} \ct^{p'}_{mj} \cC^{ni}_{mj}(p'|\mathbf{Z'}) \Big] \,,
\end{equation}
where $\mathbf{Z}=\{t,\sx,\theta,\tilde{\theta},\chi,\tilde{\chi}\}$ and $\mathbf{Z'}=\{1-t,-x,i(\thetat-\theta),i\tilde{\theta},i(\chit-\chi),i\tilde{\chi}\}$.

\subsection{Global $\cN=2$ despotic blocks}
Like in the cases described in previous sections, we would like to find the global blocks for the despotic SGCA. We again consider the case where are all the external primary fields have the equal weights. The method employed is the same, so we will just give the results. We first write the Casimirs for the algebra \eqref{SGCAdespotic}
\begin{subequations}
\begin{align}
\cC_1&=M_0^2-M_1\,M_{-1}+\frac12 H_{\frac12}\,H_{-\frac12}\,,
\\
\cC_2&=2 L_0 M_0 -\frac12 (L_1\,M_{-1}+L_{-1}\,M_1+M_1\,L_{-1}+M_{-1}\,L_1)
\nonumber\\
&\quad+\frac14 (G_{\frac12}H_{-\frac12} +H_{\frac12}G_{-\frac12}-G_{-\frac12}H_{\frac12}-H_{-\frac12}G_{\frac12})\,.
\end{align}
\end{subequations}
It can be easily seen that the eigenvalues of the Casimirs on the primary state $|\D_p,\xi_p\rangle$ are given by
\begin{equation}
\lambda_{1}^{p}=\xi_{p}^{2}\,, \qquad \lambda_{2}^{p}=\xi_{p}(2\Delta_{p} -1)\,. 
\end{equation}
The differential equations obtained by inserting the Casimir $\cC_1$  into the four point function, written in terms of the differential operators \eqref{DCas}, are given by
\begin{subequations}
\begin{align}
\cD_x^2 F_{0} & = -\frac{1}{2}t \partial_x F_{\chi \chit}\,, \\
\cD_x^2 F_{\theta \thetat} & =  -\frac{1}{2t} \xi_p^2 F_{0} + \frac12 t(2-3t) \partial_x F_{\chi \thetat}  + \frac14 (1- 2t) F_{\chi \chit} - \frac12 t (t-2) \partial_x F_{\theta \chit} \nonumber \\
&\quad  + \frac12 t (t-1) \partial_x F_{\theta \chi} + \frac{t}{2} \partial_x F_{\thetat \chit} + \frac{t}{2} F_{\theta \chi \thetat \chit} \,,  \\
\cD_x^2 F_{\chi \thetat} & = -\frac{1}{2}t(t-2) \partial_x F_{\chi \chit}\,, \\
\cD_x^2F_{\chi \chit} & =  0\,,  \\
\cD_x^2 F_{\theta \chit} & =  -\frac{1}{2}t(3t-2) \partial_x F_{\chi \chit}\,,  \\
\cD_x^2F_{\theta \chi} & =  \frac{1}{2}t\partial_{x}F_{\chi \chit}\,, \\
\cD_x^2F_{\thetat \chit} & =  \frac12 t(t-1)\partial_{x}F_{\chi \chit}\,, \\
\cD_x^2 F_{\theta \chi \thetat \chit} & =  -\frac{1}{2t} \xi_p^2 F_{\chi \chit}\,.
\end{align}
\end{subequations}
For the second Casimir the differential equations are
\begin{subequations}
\begin{align}
\cD_{tx} F_{0} & = \frac{1}{2} \left( x \partial_x F_{\chi \chit} + t(F_{\chi \thetat} + F_{\theta \chit}) \right)\,, \\
\cD_{tx}  F_{\theta \thetat} & =   \frac{1}{2t} \left[  \lambda_2^p - \frac{x}{t} \lambda_1^p \right] F_{0} + 2t(t-1) \partial_x F_{\theta \thetat} - \frac14 \left[ (2t-1) + 2t(t-2) \partial_t +4 (t-1)x \partial_x \right] F_{\theta \chit}  \nonumber \\
& \quad + \frac{x}{2} F_{\chi \chit} 
+ \frac14 \left[ (6t-1) + 2t(3t-2)\partial_t + 4(3t-1)x \partial_x \right] F_{\chi \thetat}  \nonumber \\
&\quad - \frac12 \left[ t + t(t-1) \partial_t + (2t-1)x \partial_x \right]F_{\theta \chi} - \frac{1}{2} \left[ t\partial_t + x \partial_x \right] F_{\thetat \chit} - \frac{x}{2} F_{\theta \chi \thetat \chit} \,, \\
\cD_{tx}  F_{\chi \thetat} & =   \frac{1}{2t} \lambda_1^p F_{0} + 2(t-1)t \partial_x F_{\chi \thetat} + \frac{1}{4}\left[(2t-1) + 2t(t-2) \partial_t + 4(t-1)x \partial_x \right] F_{\chi \chit}  \nonumber \\
&\quad + \frac12 t(t-1)\partial_x F_{\theta \chi} - \frac{t}{2} \partial_x F_{\thetat \chit}  + \frac{t}{2} F_{\theta \chi \thetat \chit} \,,  
\end{align}
\begin{align}
\cD_{tx} F_{\chi \chit} & =  2(t-1)t \partial_x F_{\chi \chit}\,, \\
\cD_{tx}  F_{\theta \chit} &  =   \frac{1}{2t} \lambda_1^p F_{0} + \frac14 \left[ (6t-1) + 2t(3t-2)\partial_t + 4(3t-1)x \partial_x \right] F_{\chi \chit}  + 2(t-1)t\partial_x F_{\theta \chit} \nonumber \\
&\quad - \frac12 t(t-1)\partial_x F_{\theta \chi} - \frac{t}{2} \partial_x F_{\thetat \chit}  + \frac{t}{2} F_{\theta \chi \thetat \chit} \,,  \\
\cD_{tx} F_{\theta \chi} & =  - \frac{t}{2}	\partial_{x} F_{\chi \thetat} -\frac12 \left[ t\partial_t + x \partial_x\right]F_{\chi \chit} + \frac{t}{2} \partial_x F_{\theta \chit} + t(3t-2) \partial_x F_{\theta \chi}\,, \\
\cD_{tx} F_{\thetat \chit} & =  \frac12 t(t-1)\partial_{x}F_{\chi \thetat} - \frac12 \left[ t + t(t-1) \partial_t + (2t-1)x \partial_x \right]F_{\chi \chit}  - \frac12 t(t-1)\partial_x F_{\theta \chit} \nonumber \\
& \quad + t(t-2) \partial_x F_{\thetat \chit}\,,  \\
\cD_{tx}  F_{\theta \chi \thetat \chit} & = \frac{\lambda_1^p}{2t} (F_{\chi \thetat} + \frac{x}{t} F_{\chi \chit} + F_{\theta \chit}) - \frac{\lambda_2^p }{2t} F_{\chi \chit} + 4 t(t-1) \partial_x F_{\theta \chi \thetat \chit}\,.  
\end{align}
\end{subequations}
Solving these differential equations and fixing the integration constants by comparing the small $t$ and $x$ expansion of the solution with \eqref{DESblockexp}, we can obtain the global blocks for despotic SGCA. We find:
\begin{subequations}
\label{globalDESblocks}
\begin{align}
\cA(p|\mathbf{Z}) = & \; \frac{e^{\frac{\sx}{\st}(\xi_p - 2\xi_i) } }{\st^{2\Delta_i - \Delta_p} } \frac{e^{ \frac{(1-\sqrt{1-t})}{\sqrt{1-t}} \frac{x}{t} \xi_p } }{ \sqrt{1-t}\, \left(\frac12 + \frac12 \sqrt{1-t} \, \right)^{2\Delta_p-2}}  \,,\nonumber \\
\cB(p|\mathbf{Z}) = & \; \frac{e^{\frac{\sx}{\st}(\xi_p - 2\xi_i) } }{4 \st^{2\Delta_i - \Delta_p} } \frac{e^{ \frac{(1-\sqrt{1-t})}{\sqrt{1-t}} \frac{x}{t} \xi_p } }{(1-t)\, \left(\frac12 + \frac12 \sqrt{1-t} \, \right)^{2\Delta_p}} \bigg[ \frac{x}{2\sqrt{1-t}} \theta \tilde{\theta} \\ 
\nonumber & \qquad  - (\theta + \sqrt{1-t}\,\tilde{\theta}) (\chi+ \sqrt{1-t}\, \tilde{\chi}) \bigg]\,, \\
\cC(p|\mathbf{Z}) = & \; \frac{e^{-\frac{\sx}{\st}(\xi_p + 2\xi_i) } }{ \st^{2\Delta_i + \Delta_p-1} } \frac{e^{ \frac{(1+\sqrt{1-t})}{\sqrt{1-t}} \frac{x}{t} \xi_p } }{\sqrt{1-t}} \left( 2 - 2 \sqrt{1-t}\right)^{2\Delta_p-2}\bigg[  \frac{x}{2(1-t)} \theta \tilde{\theta} \\
\nonumber & \qquad  + \frac{1}{\sqrt{1-t}} (\theta - \sqrt{1-t}\,\tilde{\theta}) (\chi - \sqrt{1-t}\, \tilde{\chi}) \bigg]\,, \\
\cD(p|\mathbf{Z}) = & \;  - \frac{1}{4\xi_p^2} \frac{e^{-\frac{\sx}{\st}(\xi_p + 2\xi_i) } }{ \st^{2\Delta_i + \Delta_p-1} } \frac{e^{ \frac{(1+\sqrt{1-t})}{\sqrt{1-t}} \frac{x}{t} \xi_p } }{\sqrt{1-t}} \left( 2 - 2 \sqrt{1-t} \right)^{2\Delta_p}\,.
\end{align}
\end{subequations}
This is the last main result of this paper: the despotic $\cN=2$ supersymmetric Galiliean conformal blocks.

\section{Conclusions}\label{se:conc}

In this paper we have set up the bootstrap program for quantum field theories with non-relativistic conformal supersymmetry. We have explicitly derived the bootstrap equations in terms of the supersymmetric Galilean conformal (SGC) blocks, which we have computed in the limit of large central charge in three different supersymmetric extensions of the Galilean conformal algebra. There are many possible paths for further research related to this work, which are needed to deepen the understanding of these types of theories. We mention a few here.

Some important aspects to analyze are the roles of null states and fusion rules in theories with SGC invariance. Generically the presence of null states imposes additional constraints on the correlation functions and on the blocks. Interestingly, in \cite{Mandal:2010gx} null states for the despotic case were found which could not be obtained from a limit of null states in the relativistic $\cN =(1,1)$ superconformal theory. It thus seems that the non-relativistic theory could have more null states than its relativistic cousin and a thorough analysis might lead to novel constraints which could be used to study the Galilean conformal blocks beyond the large $c$ limit and the non-relativistic fusion rules.

Another avenue of future research would be to compute the subleading corrections to the global blocks from suitable generalizations of \cite{Fitzpatrick:2015zha,Fitzpatrick:2015dlt} to the non-relativistic case. In that work a conformal transformation is used to map the conformal block with heavy operator insertions (with weight scaling with $c$ in the large $c$ limit) to the global block in a non-trivial background. Their result has implications for holography, where the block with both light and heavy operators corresponds to a light probe interacting with a bulk BTZ black hole. A similar setting was used in 3D flat space holography in \cite{Hijano:2018nhq} and it would be interesting to extend this to the supersymmetric Galilean conformal blocks presented here and the three dimensional flat space supergravities of \cite{Barnich:2014cwa,Lodato:2016alv}.

Concerning the application of field theory results to flat holography in three dimensions, a key issue still remains to be addressed. As mentioned in the introduction, many of the results in the flat space holography literature use the non-relativistic Galilean conformal theory and then swap the space and time coordinate with the null direction along and the spatial coordinate at $\scri$. Although this seems to work in many cases, it is still unclear how to derive these results (for our case in particular, the BMS blocks) from an intrinsically ultra-relativistic point of view. This is especially relevant if one wants to understand the BMS blocks in higher dimensions, where the isomorphism with GCA fails (in fact, in any dimension the BMS group is the conformal extension of the Carroll group \cite{Duval:2014uva}). Progress in this direction could come from a more detailed understanding of BMS invariant quantum field theories in the unitary induced representations of \cite{Barnich:2014kra}.

Further possible extensions of the present work is to include the presence of $R$-symmetry in the $\cN=2$ cases. Those theories are obtained from the non-relativistic limit of $\cN = (2,0)$ superconformal theories. In addition one could set up the same procedure in higher dimensions, where (unlike the conformal algebra) the Galilean conformal algebra remains infinite dimensional, though the isomorphism with BMS is lost. Lastly, it would be very interesting to extend the present work with a study of the implications of (a suitable limit of) modular invariance in Galilean conformal theories. This is because the necessary and sufficient conditions for a CFT to be defined consistently on two dimensional Riemann surfaces of arbitrary genus are crossing symmetry of the four-point functions on the sphere, and modular invariance of the partition function and the one-point functions on the torus \cite{Moore:1988qv}. It would be interesting to see whether the notion of non-relativistic modular invariance developed in \cite{Bagchi2013b} can similarly be used to constrain the GCA data through a non-relativistic version of the modular bootstrap \cite{Friedan:2013cba} and, in combination with crossing symmetry on the sphere, define consistent non-relativistic conformal field theories on 2 dimensional surfaces with higher genus.

\subsection*{Acknowledgements}

We are grateful to Mirah Gary for collaboration at an early stage in this project. It is a pleasure to thank A.~Bagchi, N.~Banerjee, G.~Barnich, A.~Campoleoni, H.~Gonzalez, D.~Grumiller, B.~Oblak and M.~Riegler for discussion. In addition, IL is thankful to M. Juan and M. Wurbis for illuminating interactions.
IL and WM are grateful for the hospitality at the IIT Kanpur where this project was initiated.

WM is supported by the ERC Advanced Grant “High-Spin-Grav” and by FNRS-Belgium (convention FRFC PDR T.1025.14 and convention IISN 4.4503.15).
Z  is supported by the SERB National Post Doctoral Fellowship PDF/2016/002166.

\appendix

\section{Operator product expansions}\label{sec:OPE}
In this appendix we collect some explicit expressions for the OPE and their various $\beta$ coefficients used in the main text. We will also point out some subtleties that arise when considering the supersymmetric extensions. 

\subsection{The bosonic OPE}
The operator product expansion (OPE) allows one to express the product of two primaries as a sum over primaries and its descendants. The form of the OPE is fixed by consistency with the Cartan subalgebra of the global Galilean algebra to have the form:
\begin{align}\label{GCAOPE}
\phi_m(t_m,x_m) \phi_n(t_n,x_n) = \sum_{p,\{N\}} \frac{c^{p}_{mn}}{t_{mn}^{\Delta_{mnp} - N}}e^{-\frac{x_{mn}}{t_{mn}}\xi_{mnp}} \sum_{\alpha=0}^{|l|}\beta_{mn}^{p,\{N\},\alpha} \left(\frac{x_{mn}}{t_{mn}}\right)^{\alpha} \phi_p^{\{N\}} (t_n,x_n) \;.
\end{align}
Imposing the invariance under the action of $L_0$ fixes the functional form of the OPE, while acting with $M_0$ on both sides of the OPE one can prove that $\alpha=0$ whenever no $M$-insertions are present one the right hand side. By induction one can prove that $\alpha$ must run from the 0 to the number of $M$-insertions in the descendant  $\phi_p^{\{N\}}$. Note that we are using the definitions (see also \eqref{descendants}) 
\begin{equation}
\phi_p^{\{N\}} (t_n,x_n) = L_{-\{k\}}M_{-\{l\}} \phi_p(t_n,x_n) =  L_{-k_1} \ldots L_{-k_i} M_{-l_i} \ldots M_{-l_j} \phi_p(t_n,x_n)\,.
\end{equation}
The coefficients $\beta_{mn}^{p,\{N\},\alpha}$ where computed in \cite{Bagchi:2017cpu} by deriving recursive relations between the states. They can also be computed by acting with the raising operators $L_n$ and $M_n$ as differential operators \eqref{GCAcommutators} on the three point functions and comparing like powers of $x$:
\begin{align}\label{betas}
& \sum_{\alpha = 0}^{|l|} \beta_{mn}^{\,p,\{N\},\alpha} x^{\alpha}  =  \sum_{\{N'\}} \frac{e^{\,x\, \xi_{mnp}} }{c^p_{mn}} \mathfrak{M}^{\{N\},\{N'\}}  \bra{\Delta_p,\xi_p,\{N'\}} \phi_m(t,x) \ket{\Delta_n,\xi_n} \Big|_{t = 1}\,. 
\end{align}
The explicit expressions for the OPE coefficients are given in \cite{Bagchi:2017cpu} up to level 2.\footnote{Note that we are using slightly different conventions here. To recover the results in \cite{Bagchi:2017cpu} one should take $u \to t$ and $v \to -x$.} \\

\subsection{$\cN=1$} \label{se:C.2}
The operator product expansion for SGCA primary fields can be derived by requiring it to have the proper scaling under the Cartan subalgebra of the $\cN=1$ SGCA global algebra \eqref{SGCAfieldcom}. The most general OPE reads: 
\begin{align}\label{SGCAOPE}
\Phi_m(t_m,x_m,\theta_m) \Phi_n(t_n,x_n,\theta_n) =& \sum_p \frac{c^p_{mn}}{t_{mn}^{\Delta_{mnp}}}e^{-\frac{{\bf x}_{mn}}{t_{mn}}\xi_{mnp}} \sum_{\{N\},\alpha,\gamma} \beta_{mn}^{p,\{N\},\alpha,\gamma}\, t_{mn}^{N-\alpha-\frac{q}{2}-\gamma}\, {\bf x}_{mn}^{\alpha} 
\nonumber\\
&\qquad\qquad \times(\theta_m-\theta_n)^{\,q}\;(\theta_m\theta_n)^\gamma\; \Phi_p^{\{N\}} (t_n,x_n,\theta_n)\,,
\end{align}
where $q=1$ and $\gamma =0$ whenever $N \in \mathbb{Z} + \frac12$ and $q=0\,, \;\gamma=0,1$ when $N \in \mathbb{Z}$. 
Clearly, when the second set of coordinates is fixed to be the origin of superspace, $(t_n,x_n,\theta_n)=(0,0,0)$ the OPE simplifies considerably since ${\bf x}_{mn}\to x$, $t_{mn}\to t$ and $\gamma\to 0$:
\begin{equation}\label{SGCAOPE2}
\Phi_m(t,x,\theta) \Phi_n(0,0,0) = \sum_p \frac{c^p_{mn}}{t^{\Delta_{mnp}}}e^{-\frac{ x}{t}\xi_{mnp}} \sum_{\{N\},\alpha} \beta_{mn}^{p,\{N\},\alpha,0}\, t^{N-\alpha-\frac{q}{2}}\, x^{\alpha} \theta^{\,q}\;\; \Phi_p^{\{N\}} (0,0,0)\,.
\end{equation}
Here we will be treating the more general case to explicitly show some of the subtleties that arise due to a non-zero $\theta_n$. This is needed if we wish to express the blocks \eqref{N=1block} in terms of OPE coefficients.

Let us first look at the OPE coefficients $\beta_{mn}^{p,\{N\},\alpha,0}$. They can be computed by acting with the differential operators \eqref{SGCAfieldcom} on the three point function and comparing like coefficients of $x$ and $\theta$ 
\begin{align}
\label{eq:op_betaN1}
& \sum_\alpha \beta_{mn}^{p,\{N\},\alpha,0} x^\alpha (-\theta)^q = \sum_{\{N'\}} \frac{e^{ x\xi_{mnp}}}{c^p_{mn}} \mathfrak{M}^{\{N\},\{N'\}} \bra{\Delta_p,\xi_p,\{N'\}} \Phi_m(t,x,\theta) \ket{\Delta_n,\xi_n} \bigg|_{t=1}\,. 
\end{align}
Specifically this implies that the three point function with descendants at $(t,x,\theta) = (1,0,0)$ is simply
\begin{equation}\label{beta0s}
\bra{\Delta_p, \xi_p, \{N\}} \Phi_m (1,0,0) \ket{\Delta_n,\xi_n} = \sum_{\substack{\{N'\} \\ |N'|\in \mathbb{Z}}} c^p_{mn}\beta_{mn}^{p, \{N'\},0,0} \mathfrak{M}_{\{N'\},\{N\}}\,,
\end{equation}
where the sum on the right hand side only runs over integer levels. This accounts for the first three point function in \eqref{N=1block}.

The OPE coefficients are related to one another according to recursion relations, which are obtained by acting with lowering operators on both sides of the OPE \eqref{SGCAOPE2} 
\begin{subequations}
\begin{align}
L_k\ket{N+k,\alpha,q}&=\Big(\Delta_p-\Delta_n+N
+k(\alpha+\tfrac{q}2+\Delta_m)\Big)\,\ket{N,\alpha,q} \\
& \qquad \qquad
+k(\xi_m\,k+\xi_p-\xi_n)\ket{N,\alpha-1,q}\;,
\nonumber\\
M_k\,\ket{N+k,\alpha,q}&=(\alpha+1)\ket{N,\alpha+1,q}+(\xi_m\,k+\xi_p-\xi_n)\ket{N,\alpha,q}\;,
\\
Q_r\ket{N+r,\alpha,0}&=\ket{N,\alpha,1}\;,
\\
Q_r\ket{N+r,\alpha,1}&=\tfrac{1}{2} \Big((\alpha+1)\ket{N,\alpha+1,0}+(2\,\xi_m\, r+\xi_p-\xi_n)\ket{N,\alpha,0}\Big)\;,
\end{align}
\end{subequations}
with $k,r>0$. The states $\ket{N,\alpha,q}$ are defined as:
\begin{equation}
\label{eq:statesN1}
\ket{N,\alpha,q}=\sum_{\stackrel{\{\vec{k},\vec{l},\vec{r}\}}{|\vec{k}|+|\vec{l}|+|\vec{r}|=N}}\beta^{p,\{\vec{k},\vec{l},\vec{r}\},\alpha,0}_{mn}\,L_{-\vec{k}} \; M_{-\vec{l}}  \; Q_{-\vec{r}}\; \ket{\Delta_p,\xi_p}\;,
\end{equation} 
with $ \beta^{\{0,0,0\},0,0}=1$ and $q=(0,1)$ when $N \in (\mathbb{Z},\mathbb{Z}+\frac12)$ respectively. 
It is easy to check that combining both the recursion relations involving the supergenerator $Q_{r}$ one obtains the recursion relation imposed by $\frac12 M_{2r}$.

The first few coefficients of the $\cN=1$ supersymmetric GCA$_2$ OPE \eqref{SGCAOPE2} are
\begin{subequations}
\begin{align}
\label{eq:level1/2}
\text{level } \tfrac12: && \beta_{mn}^{p,\{0,0,\tfrac12\},0,0} & =\frac{\xi_{pmn}}{2\,\xi_p}\,, \\
\text{level } 1: && \beta_{mn}^{p,\{1,0,0\},0,0} & = \beta_{mn}^{p\{0,1,0\},1}= \frac{\xi_{pmn}}{2\xi_p} \,, \\
&& \beta_{mn}^{p,\{0,1,0\},0,0} & = \frac{\Delta_{pmn}}{2\xi_p}-\frac{\xi_{pmn}\Delta_p}{2\xi_p^2} \,, \\
\text{level } \tfrac32: &&
\beta_{mn}^{p,\{0,1,\tfrac12\},1,0} & = \beta_{mn}^{p,\{1,0,\tfrac12\},0}=\frac{\xi_{pmn}^2}{4\,\xi_p^2} \,, \\
&& 
\beta_{mn}^{p,\{0,0,\tfrac32\},0,0} & = \frac32\,\frac{1}{c_M+3\,\xi_p}\Big(\xi_m+ \xi_n-\frac{(\xi_m-\xi_n)^2}{\xi_p}\Big) \,,
\\
&& 
\beta_{mn}^{p,\{0,1,\tfrac12\},0,0} & = \frac{1}{8\xi_p^3}\Big[\Big(2\xi_p\Delta_{pmn}-2\Delta_p\xi_{pmn}+\xi_{pnm}\Big)\xi_{pmn} 
\nonumber\\
&& &\qquad\qquad-\frac{12\,\xi^2_p}{c_M+3\xi_p}\Big(\xi_m+\xi_n-\frac{(\xi_m-\xi_n)^2}{\xi_p}\Big)\Big] \,.
\end{align}
\end{subequations}
To obtain the second three point function in \eqref{N=1block}, we need the coefficients $\beta_{mn}^{p,\{\vec{k},\vec{l},\vec{r}\},\alpha,1}$. One can find the first few by acting with the raising operator $Q_{-\frac12}$ on both sides of the OPE (once as a generator and once as a differential operator) and compare like powers of the coordinates. We obtain
\begin{equation}
\beta_{mn}^{p,\{0,0,0\},0,1}=\beta_{mn}^{p,\{1,0,0\},0,1}=\beta_{mn}^{p,\{0,1,0\},1,1}=0\;, \quad \beta_{mn}^{p,\{0,1,0\},0,1}=\beta_{mn}^{p,\{0,0,\tfrac12\},0,0}\,.
\end{equation}
A formula which fixes the coefficients $\beta_{mn}^{p, \{N\},\alpha,1}$ along the lines of \eqref{eq:op_betaN1} exists, but is more complicated due to the presence of the Grassmann variables in the second operator $\Phi_n$. These coefficients are computed by computing the $\theta \eta$-component of the three point function
\begin{equation}\label{N=1tp}
(c_{mn}^p)^{-1} e^{\sx\, \xi_{mnp}} \bra{\Delta_p,\xi_p, \{N'\}} \Phi_m (t,x,\theta) \Phi_n(0,0,\eta) \ket{0} |_{t\to 1} \,,
\end{equation}
in two different ways. First one can compute it by acting with the differential operator form of the descendants \eqref{SGCAfieldcom} in $\{N'\}$ on the three point function $\bra{\Delta_p,\xi_p} \Phi_m (t,x,\theta) \Phi_n(0,0,\eta) \ket{0} = e^{- \frac{\sx}{t}\xi_{mnp}}t^{-\Delta_{mnp}}$ and then read off the $\theta \eta$ component.
	
This should then equal all possible $\theta \eta$ components which appear when expanding the primaries $\Phi_m \Phi_n$  with the OPE. Looking closely at \eqref{SGCAOPE} we see that there are three possible sources for these terms. One are the $\gamma=1$ terms, which we are after. Another comes from $\sx^\alpha$ which expanded in Grassmann coordinates reads $x^\alpha - \frac{\alpha}{2} x^{\alpha-1} \theta \eta$. 
The third contribution comes from the fact that $\bra{\Delta_p,\xi_p,\{N'\}}\Phi_p^{\{N\}}(0,0,\eta)\ket{0}$ is not the diagonal Kac matrix, but receives contributions from half integer level descendants in $\{N\}$ times $\eta$. Explicitly, expanding $\Phi_p(0,0,\eta)\ket{0} = \ket{\Delta_p,\xi_p} + \eta Q_{-1/2} \ket{\Delta_p,\xi_p}$ we have that
\begin{align}
\bra{\Delta_p,\xi_p,\{N'\}}\Phi_p^{\{N\}}(0,0,\eta)\ket{0}  = & \; \mathfrak{M}_{\{N'\},\{N\}} \\ 
&  - \eta \bra{\Delta_p,\xi_p,\{N'\}} L_{-\{k\}} M_{-\{l\}} Q_{-\{r\}} Q_{-1/2} \ket{\Delta_p,\xi_p}\,. \nonumber
\end{align}
Putting this all together, we see that the $\theta \eta$ terms of \eqref{N=1tp} should equal
\begin{align}\label{threebetas}
\sum_{\substack{\{N\},\alpha\\ |N| \in \mathbb{Z}}} \left( \beta_{mn}^{p,\{N\},\alpha,1} x^\alpha - \beta_{mn}^{p,\{N\},\alpha,0}\frac{\alpha}{2} x^{\alpha-1} \right) &\mathfrak{M}_{\{N'\},\{N\}} \nonumber  \\
&  - \sum_{\substack{\{N\},\alpha\\ |N| \in \mathbb{Z}+\frac12}} \beta_{mn}^{p,\{N\},\alpha,0} x^\alpha \mathfrak{M}_{\{N'\},\{N+\frac12\}} \,,
\end{align}
where here $\{N+\frac12\}$ denotes the descendant generated by $ L_{-\{k\}} M_{-\{l\}} Q_{-\{r\}} Q_{-1/2} $. This fixes the $\beta$ coefficients with $\gamma=1$ in terms of known $\beta$'s. 
In general these relations could become quite cumbersome to solve, however there is an argument which allows us to find the contribution to the blocks \eqref{N=1block} in the limit of large central charge. In this limit the only descendants contributing to the blocks are those of the global subalgebra $\{L_{-1}, M_{-1}, Q_{-1/2}\}$. When computing the $\theta \eta$ contributions to \eqref{N=1tp} for these descendants by acting on the three point function with the differential operators \eqref{SGCAfieldcom} we see that it vanishes. This implies that \eqref{threebetas} should vanish as well. Hence all coefficients $\beta_{mn}^{p,\{N\},\alpha,1}$ are such that they cancel the other contributions to \eqref{threebetas} in the limit of large central charges. This implies that all $\theta \eta$ dependence in the second three point function of the block \eqref{N=1block} is captured by the exponent $e^{-\frac{\sx}{t}\xi_{mnp}}$ from the OPE \eqref{SGCAOPE}. So we may write
\begin{align}
& \lim_{c_{M/L}\to\infty} \bra{\Delta_p,\xi_p, \{N'\}} \Phi_m(t,x,\theta) \Phi_n(0,0,\eta)\ket{0} \nonumber \\ 
& \qquad \qquad \qquad \qquad \qquad \qquad \qquad = \frac{c^p_{mn}e^{-\frac{\sx}{t}\xi_{mnp}}}{t^{\Delta_{mnp}} }\sum_{\{G\},\alpha}  \beta_{mn}^{p,\{G\},\alpha,0} t^{G-\alpha} x^\alpha \mathfrak{M}_{\{G'\},\{G\}}\,,
\end{align}
where now $\{G\}$ denotes only descendants in the global subalgebra at level $G$, such that $\ket{\Delta_p,\xi_p,\{G\}} = (L_{-1})^k (M_{-1})^l (Q_{-1/2})^q \ket{\Delta_p,\xi_p}$ and $G = k+l+q/2$ and $q=(0,1)$.

This implies that the global $\cN=1$ supersymmetric Galilean conformal blocks $g^{ji}_{mn}$ can be expressed in terms of the OPE coefficients as
\begin{align}\label{SGCAblockN1}
g^{ji}_{mn}(p|t,x,\theta,\eta) & 
=  \frac{ e^{-\xi_{mnp}\frac{\mathbf{x}}{t}} }{t^{\Delta_{mnp}}}\sum_{\substack{\{G\},\{G'\},\alpha \\ |G'|=|G| \in \mathbb{Z}}} t^{G-\alpha}x^\alpha  \beta_{ji}^{p,\{G'\},0,0}\, \mathfrak{M}_{ \{G'\}, \{G\} } \beta_{mn}^{p,\{G\},\alpha,0} \,.
\end{align}
Note that the factor of $x^\alpha$ appearing here is not the supersymmetric coordinate $\sx^\alpha$, as we have shown that all the $\theta \eta$ dependent terms are in the exponent.

\subsection{$\cN=2$ Democratic}
In the $\cN=2$ cases the appearance of a second structure constant in the three point function implies that the OPE will also receive two separate contributions, leading to two sectors with their own OPE coefficients, an `untilded sector' with coefficients  $\beta_{mn}^{p,\{N\},\alpha}$ and a `tilded sector' denoted by $\bt_{mn}^{p,\{N\},\alpha}$:
\begin{align}
\label{demSGCAalgebra}
& \Phi_m(t,x,\theta^{\pm}) \Phi_n(0,0,0,0) = \, \sum_{p,\{N\},\alpha} e^{-\frac{x}{t}\xi_{mnp}}t^{-\Delta_{mnp}+N-\alpha} x^{\alpha}\Big( c^p_{mn} \beta_{mn}^{p,\{N\},\alpha}t^{-\frac{R+S}{2}} (\theta^+)^R(\theta^-)^S \nonumber \\ 
& \qquad + \ct^p_{mn} \bt_{mn}^{p,\{N\},\alpha}t^{\frac{R+S-2}{2}} (\theta^+)^{1-R}(\theta^-)^{1-S} \Big) L_{-\{k\}} M_{-\{l\}} Q^+_{-\{r\}} Q^-_{-\{s\}}\Phi_p (0,0,0,0)\,.
\end{align}
Here $R$ and $S$ are either 0 or 1, depending on the number of $Q^{\pm}$ descendants: an even number of $Q^+$ operators in the descendant implies $|r| \in \bZ$ and $R=0$, while an odd number implies $|r| \in \bZ +\tfrac12$ and $R=1 $. Likewise for $S$ and the number of $Q^-$ operators. 

The general expressions for the OPE with non-zero Grassmann coordinates on $\Phi_n$, say $\eta^\pm$, is clearly much more involved than in the $\cN=1$ case. This time we would need to consider quadratic and quartic combinations of the four Grassmann variables $\theta^\pm,\eta^\pm$. However, since the three-point functions relevant for the SGC blocks (i.e. with non-zero fermionic coordinate dependence) can also be computed by acting with the differential operators \eqref{demSGCAfieldcom} on the three point function of primaries we will focus only on finding the coefficients $ \beta_{mn}^{p,\{N\},\alpha} $ and $ \bt_{mn}^{p,\{N\},\alpha} $ for the OPE \eqref{demSGCAalgebra}.
These coefficients can be computed by acting on the three point function with the differential operators \eqref{demSGCAfieldcom} and comparing like powers of $x$ and $\theta^{\pm}$:
\begin{align}
& \sum_\alpha  (-)^{R+S} x^{\alpha}\Big( c^p_{mn} \beta_{mn}^{p,\{N\},\alpha} (\theta^+)^R(\theta^-)^S + \ct^p_{mn} \bt_{mn}^{p,\{N\},\alpha} (\theta^+)^{1-R}(\theta^-)^{1-S} \Big) =  \nonumber \\
& \qquad  \sum_{\{N'\}} e^{x \xi_{mnp}} \mathfrak{M}^{\{N\},\{N'\}} \delta_{Q^-_{+\{s\}}} \delta_{Q^+_{+\{r\}}}\delta_{M_{+\{l\}}}\delta_{L_{+\{k\}}} \bra{\Delta_p,\xi_p} \Phi(t,x,\theta^{\pm}) \ket{\Delta_n,\xi_n} \Big|_{t=1}\,.
\end{align}
The coefficients $\beta$ and $\bt$ can alternatively be found by solving a set of recursive relations. First we can split the OPE into two parts explicitly, one for each independent structure constant, by writing 
\begin{align}\label{OPEstates}
\Phi_m(t,x,\theta^{\pm})\ket{\Delta_n, \xi_n} = \sum_{p, \{N\}, \alpha} & e^{-\frac{x}{t}\xi_{mnp} } t^{-\Delta_{mnp} + N - \alpha} x^{\alpha} \Big[  c^p_{mn} t^{-\frac{R+S}{2} } (\theta^+)^R (\theta^-)^S \ket{N,\alpha,R,S} \nonumber \\  & 
+ \ct^p_{mn} t^{\frac{R+S-2}{2}}(\theta^+)^{1-R} (\theta^-)^{1-S} \widetilde{\ket{N,\alpha,R,S}} \Big]\,,
\end{align}
where we have defined the states
\begin{subequations}
\begin{equation}
\ket{N,\alpha,R,S} = \sum_{\substack{\{\vec{k},\vec{l},\vec{r},\vec{s}\}\\|\vec{k}|+|\vec{l}|+|\vec{r}|+|\vec{s}|=N} }
\beta_{mn}^{p,\{\vec{k},\vec{l},\vec{r},\vec{s}\},\alpha} L_{-\vec{k} } \; M_{-\vec{l}} \; Q^+_{-\vec{r}} \; Q^-_{-\vec{s}}\;  \ket{\Delta_p,\xi_p}\,. 
\end{equation}
and
\begin{equation}
\widetilde{\ket{N,\alpha,R,S}} = \sum_{\substack{\{\vec{k},\vec{l},\vec{r},\vec{s}\}\\|\vec{k}|+|\vec{l}|+|\vec{r}|+|\vec{s}|=N} } \bt_{mn}^{p,\{\vec{k},\vec{l},\vec{r},\vec{s}\},\alpha} L_{-\vec{k} } \; M_{-\vec{l}} \; Q^+_{-\vec{r}} \; Q^-_{-\vec{s}} \; \ket{\Delta_p,\xi_p}\,.
\end{equation}
\end{subequations}
Here $R \; (S)$ is 0 for an even and 1 for an odd number of $Q^+ \; (Q^-)$ descendants. The recursive relations are obtained by acting with raising operators on both sides of \eqref{OPEstates} and comparing like powers of the coordinates. In the case for the democratic SGCA the recursive relations will not mix the contributions from $\beta$ coefficients with those from $\bt$ coefficients and so we may write the recursive relations separately for the two sectors. They are
\begin{subequations}
\begin{align}
L_k& \ket{N+k,\alpha,R,S} = \Big(\Delta_p-\Delta_n+N
+k(\alpha+\tfrac{R+S}{2}+\Delta_m)\Big)\,\ket{N,\alpha,R,S}
\\ &  +k(\xi_m\,k+\xi_p-\xi_n)\ket{N,\alpha-1,R,S}\;,
\nonumber\\
M_k& \ket{N+k,\alpha,R,S} =(\alpha+1)\ket{N,\alpha+1}+(\xi_m\,k+\xi_p-\xi_n)\ket{N,\alpha,R,S}\;,
\\
Q^+_r & \ket{N+r,\alpha,R,S} = (-)^S (1-R)\ket{N,\alpha,1-R,S}   \\&
+ \tfrac{(-)^S R}{2} \Big((\alpha+1)\ket{N,\alpha+1,1-R,S}+(2\,\xi_mr+\xi_p-\xi_n)\ket{N,\alpha,1-R,S}\Big)\;,
\nonumber\\
Q^-_r & \ket{N+r,\alpha,R,S} = (1-S)\ket{N,\alpha,R,1-S}    \\&
+ \tfrac{S}{2} \Big((\alpha+1)\ket{N,\alpha+1,R,1-S}+(2\,\xi_mr+\xi_p-\xi_n)\ket{N,\alpha,R,1-S}\Big)\;, \nonumber
\end{align}	
\end{subequations}	
and
\begin{subequations}
\begin{align}
L_k & \widetilde{\ket{N+k,\alpha,R,S}} = \Big(\Delta_p-\Delta_n+N
+k(\alpha-\tfrac{R+S-2}{2}+\Delta_m)\Big)\,\widetilde{\ket{N,\alpha,R,S}} \\ & 
+k(\xi_m\,k+\xi_p-\xi_n)\widetilde{\ket{N,\alpha-1,R,S}}\;,
\nonumber\\
M_k & \widetilde{\ket{N+k,\alpha,R,S}} = (\alpha+1)\widetilde{\ket{N,\alpha+1}}+(\xi_m\,k+\xi_p-\xi_n)\widetilde{\ket{N,\alpha,R,S}}\;,
\\
Q^+_r & \widetilde{\ket{N+r,\alpha,R,S}} =(-)^{S+1} R \widetilde{\ket{N,\alpha,1-R,S}}  \\
& + \tfrac{(-)^{S+1}(1-R)}{2} \Big((\alpha+1 )\widetilde{\ket{N,\alpha+1,1-R,S}}+(2\,\xi_mr+\xi_p-\xi_n) \widetilde{\ket{N,\alpha,1-R,S}}\Big)\;,
\nonumber\\
Q^-_r & \widetilde{\ket{N+r,\alpha,R,S}} = S\widetilde{\ket{N,\alpha,R,1-S}}  \\ &
+ \tfrac{1-S}{2} \Big((\alpha+1)\widetilde{\ket{N,\alpha+1,R,1-S}}+(2\,\xi_mr+\xi_p-\xi_n)\widetilde{\ket{N,\alpha,R,1-S}}\Big)\;, \nonumber
\end{align}
\end{subequations}
for $k,r>0$. 

The beta coefficients up to level 1 are given by:
\begin{subequations}
\begin{align}
\label{eq:level1/2N2}
\text{level } \tfrac12: && \beta_{mn}^{p,\{0,0,\tfrac12,0\},0} & =\beta_{mn}^{p,\{0,0,0,\tfrac12\},0}=\frac{\xi_{pmn}}{2\,\xi_p}\,,
 \\
&& \tilde{\beta}_{mn}^{p,\{0,0,\tfrac12,0\},0} & =-\tilde{\beta}_{mn}^{p,\{0,0,0,\tfrac12\},0}=-\frac{1}{\xi_p}\,,
\\
\text{level } 1: && \beta_{mn}^{p,\{1,0,0,0\},0} & = \beta_{mn}^{p\{0,1,0,0\},1}= \frac{\xi_{pmn}}{2\xi_p} \,,
\\
&& \beta_{mn}^{p,\{0,1,0,0\},0} & = \frac{\Delta_{pmn}}{2\xi_p}-\frac{\xi_{pmn}\Delta_p}{2\xi_p^2} \,,
\\
&& \tilde{\beta}_{mn}^{p,\{1,0,0,0\},0} & = \tilde\beta_{mn}^{p\{0,1,0,0\},1}= \frac{\xi_{pmn}}{2\xi_p} \,,
\end{align}
\begin{align}
&& \tilde\beta_{mn}^{p,\{0,1,0,0\},0} & = \frac{\Delta_{pmn}+1}{2\xi_p}-\frac{\xi_{pmn}\Delta_p}{2\xi_p^2} \,,
\\
&&\beta_{mn}^{p,\{0,0,\frac12,\frac12\},0} & =-\frac{\xi_{pmn}^2}{4\xi^2_p} \,,
\\
&&\tilde\beta_{mn}^{p,\{0,0,\frac12,\frac12\},0} & =\frac{1}{\xi^2_p} \,.
\end{align}
\end{subequations}

\subsection{$\cN=2$ Despotic}
For the despotic algebra \eqref{SGCAdespotic}, the functional form of the operator product expansion is identical to the democratic one \eqref{demSGCAalgebra} up to renaming variables. This is due to the fact that $L_0$ has not changed with respect to the democratic case (while taking $\theta^+ \to \theta $ and $\theta^- \to \chi$). However, the $M_0$ action now receives contributions from fermionic generators. In particular, it contains a term $\frac12 \theta \partial_{\chi}$ whose action on the OPE is able to mix what we called the tilded and untilded sector in the last subsection. In general, we can write the OPE as
\begin{align}
\label{despSGCAalgebra}
& \Phi_m(t,x,\theta,\chi) \Phi_n(0,0,0,0) = \, \sum_{p,\{N\},\alpha} e^{-\frac{x}{t}\xi_{mnp}}t^{-\Delta_{mnp}+N-\alpha} x^{\alpha}\Big( \gamma_{mn}^{p,\{N\},\alpha}t^{-\frac{R+S}{2}} (\theta)^R(\chi)^S \nonumber \\ 
& \qquad + \tilde{\gamma}_{mn}^{p,\{N\},\alpha}t^{\frac{R+S-2}{2}} (\theta)^{1-R}(\chi)^{1-S} \Big) L_{-\{k\}} M_{-\{l\}} G_{-\{r\}} H_{-\{s\}}\Phi_p (0,0,0,0)\,.
\end{align}
Here $R$ ($S$) is 0 for an odd number and 1 for an even number of $G_r$ ($H_r$) descendants and we have introduced the OPE coefficients $\gamma_{mn}^{p,\{N\},\alpha}$ and $ \tilde{\gamma}_{mn}^{p,\{N\},\alpha}$. By comparing with the three point function we find $\gamma_{mn}^{p,\{\vec{0}\},0} =c_{mn}^p$ and $\tilde{\gamma}_{mn}^{p,\{\vec{0}\},0} = \ct_{mn}^{p}$, but unlike in the democratic case, at higher levels the coefficients $\gamma_{mn}^{p,\{N\},\alpha} $ do not correspond solely to terms related to $c_{mn}^p$ and likewise for $\tilde{\gamma}$ and $\ct$. Another novel feature is that now the range of $\alpha$ is runs from 0 to the number of $M$ insertions in the descendant plus 1 if the number of $H$ insertions is odd at half integer level. 

Generically, the OPE coefficients can be computed in a similar fashion as the last subsections, now by acting on the three point function \eqref{threeptcctDES} with the differential operators \eqref{despSGCAfieldcom} and comparing like powers of $x$, $\theta$ and $\chi$
\begin{align}
& \sum_\alpha  (-)^{R+S} x^{\alpha}\Big( \gamma_{mn}^{p,\{N\},\alpha} \theta^R\,\chi^S +  \tilde{\gamma}_{mn}^{p,\{N\},\alpha} \theta^{1-R}\,\chi^{1-S}  \Big) =   \\ \nonumber
& \quad  \sum_{\{N'\}} e^{x \xi_{mnp}} (\mathfrak{M}_{\{N\},\{N'\}})^{-1} \delta_{H_{+\{s\}}} \delta_{G_{+\{r\}}}\delta_{M_{+\{l\}}}\delta_{L_{+\{k\}}} \bra{\Delta_p,\xi_p} \Phi(t,x,\theta,\chi) \ket{\Delta_n,\xi_n}\Big|_{t=1}\,.
\end{align}
To find the dependency of the OPE coefficients on solely $c_{mn}^p$ (or $\ct_{mn}^p$) one can simply replace the three point function above by $\vev{\Phi_p \Phi_m \Phi_n}_{c (\ct)}$ defined in \eqref{threeptcctDES}. 

The result for the $\gamma$ coefficients up to level 1 reads:
\begin{subequations}
	\label{eq:level1N2desp}
\begin{align}
\text{level } \tfrac12: && \gamma_{mn}^{p,\{0,0,\tfrac12,0\},0} & = \frac{\xi_{pmn}}{2\,\xi_p} c^p_{mn} + \frac{\ct_{mn}^p}{2\xi_p}\,,
\\
&& \gamma_{mn}^{p,\{0,0,0,\tfrac12\},0}& = \frac{\xi_{pmn}}{2\,\xi_p} c^p_{mn} - \frac{\ct_{mn}^p}{2\xi_p}\,,
\\
&& \tilde{\gamma}_{mn}^{p,\{0,0,0,\tfrac12\},0}&=  \frac{c_{mn}^p ((\Delta_{m} -\Delta_n) \xi_p - (\xi_m-\xi_n)\Delta_p ) - \ct_{mn}^p \Delta_p}{2\xi^2_p} \,,
\\
&& \tilde{\gamma}_{mn}^{p, \{0,0,0,\tfrac12\},1} & = \frac{\ct_{mn}^p}{2\xi_p} \,, \\
\text{level } 1: && \gamma_{mn}^{p,\{1,0,0,0\},0} & = \gamma_{mn}^{p\{0,1,0,0\},1}= \frac{\xi_{pmn}}{2\xi_p} c_{mn}^p \,,
\\
&& \gamma_{mn}^{p,\{0,1,0,0\},0} & = \frac{2((\Delta_m-\Delta_n)\xi_p-(\xi_m-\xi_n)\Delta_p)c_{mn}^p + \ct_{mn}^p}{4\xi^2_p} \,,
\\
&& \tilde{\gamma}_{mn}^{p,\{1,0,0,0\},0} & = \tilde{\gamma}_{mn}^{p\{0,1,0,0\},1}= \frac{\xi_{pmn}}{2\xi_p} \ct_{mn}^p \,,
\\
&& \hat\gamma_{mn}^{p,\{0,1,0,0\},0} & = \frac{2((1+\Delta_m-\Delta_n)\xi_p-(\xi_m-\xi_n)\Delta_p)\ct_{mn}^p + c_{mn}^p \xi_{pmn}^2}{4\xi^2_p} \,,
\\
&&\gamma_{mn}^{p,\{0,0,\frac12,\frac12\},0} & = - \frac{\xi_{pmn}^2}{4\xi^2_p} c_{mn}^p \,,
\\
&&\tilde{\gamma}_{mn}^{p,\{0,0,\frac12,\frac12\},0} & = - \frac{1}{4\xi^2_p} \ct_{mn}^p \,.
\end{align}
\end{subequations}
And all coefficients up to level 1 which are not listed here vanish. The derivation of these coefficients by recursive relations is complicated by the fact that the coefficient mix among the two sectors. We hence refrain from listing the recursive relations for the despotic case here.

\providecommand{\href}[2]{#2}\begingroup\raggedright\endgroup

\end{document}